\documentclass[useAMS,usenatbib]{mn2e}
\usepackage{times}

\usepackage{amssymb}
\usepackage{psfig}

\title[Bolometric correction of cool stars]
{Bolometric correction and spectral energy distribution of cool stars in
Galactic clusters\thanks{Based on observations made at La Palma, at the Spanish Observatorio del
Roque de los Muchachos of the IAC, with the Italian Telescopio Nazionale Galileo (TNG) 
operated by the Fundaci\'on Galileo Galilei of INAF.}}
\author[A. Buzzoni et al.]{A. Buzzoni$^1$, L. Patelli$^1$, M. Bellazzini$^1$, 
F. Fusi Pecci$^1$, \& E. Oliva$^{2}$\\
$^1$ INAF - Osservatorio Astronomico di Bologna, Via Ranzani 1, 40127 Bologna, Italy;\\
$^2$ INAF - Osservatorio Astrofisico di Arcetri, L.go E. Fermi 5, 50125 Firenze, Italy\\
\\
}

\begin{document}

\date{Accepted ... Received ... in original form}

\pagerange{\pageref{firstpage}--\pageref{lastpage}} \pubyear{2006}

\maketitle

\label{firstpage}

\begin{abstract}

In this work we have investigated the relevant trend of the bolometric
correction  (BC) at the cool-temperature regime of red giant stars and its
possible dependence  on stellar metallicity. Our analysis relies on a wide
sample of optical-infrared spectroscopic observations, along the
3500~\AA~$\Rightarrow 2.5\mu$m wavelength range, for a grid of 92 red giant
stars in five (3 globular + 2 open) Galactic clusters, along the full
metallicity range covered by the bulk of the stars, $-2.2 \le [Fe/H] \le +0.4$.

Synthetic $BVR_cI_cJHK$ photometry from the derived spectral energy
distributions allowed us to obtain robust temperature (T$_{\rm eff}$)  estimates
for each star, within $\pm 100$~K or less. According to the appropriate temperature
estimate, black-body extrapolation of the observed spectral energy distribution (SED)
allowed us to assess the unsampled flux beyond the wavelength limits of our survey.
For the bulk of our red giants, this fraction amounted to 15\% of the total bolometric
luminosity, a figure that raises up to 30\% for the coolest targets 
(T$_{\rm eff} \lesssim 3500$~K). Allover, we trust to infer stellar M$_{\rm bol}$
values with an internal accuracy of a few percent. Even neglecting any correction
for lost luminosity etc. we would be overestimating M$_{\rm bol}$ by $\la 0.3$ mag, 
in the worst cases. Making use of our new database, we provide a set of fitting functions for
the V and K BC vs.\ T$_{\rm eff}$ and vs. $(B-V)$ and $(V-K)$ broad-band colors,
valid over  the interval $3300~{\rm K} \le {\rm T}_{\rm eff} \le 5000$~K,
especially suited for Red Giants.

The analysis of the BC$_V$ and BC$_K$ estimates along the wide range of
metallicity spanned by our stellar sample show no evident drift with [Fe/H].
Things may be different for the B-band correction, where the blanketing effects
are more and more severe. A drift of $\Delta (B-V)$ vs. [Fe/H] is in fact
clearly  evident from our data, with metal-poor stars displaying a ``bluer''
$(B-V)$ with  respect to the metal-rich sample, for fixed T$_{\rm eff}$.

Our empirical bolometric corrections are in good overall agreement with most of
the existing theoretical and observational determinations, supporting the
conclusion that (a) $BC_K$ from the most recent studies are reliable within $\la
\pm 0.1$ over the whole color/temperature range considered in this paper, and
(b) the same conclusion apply to $BC_V$ only for stars warmer than  $\simeq
3800$~K. At cooler temperatures the agreement is less general, and  MARCS models
are the only ones providing a satisfactory match to observations, in particular
in the $BC_V$ vs.\ $(B-V)$ plane.
\end{abstract} 


\section{Introduction}

A physical assessment of the bolometric emission of stars is a mandatory
step for any attempt to self-consistently link observations and 
theoretical predictions of stellar evolution. The importance of 
this comparison actually reverberates into a wide range of primary 
astrophysical questions, ranging from the validation of the reference 
input physics for nuclear reactions in the stellar interiors to 
the study of integrated spectrophotometric properties of distant galaxies,
through stellar population synthesis models.

By definition, the effective temperature (T$_{\rm eff}$) and physical size ($R$)
of a star provide the natural constraint to its emerging flux, as
$L \propto R^2\,T_{\rm eff}^4$.
If $L$ is a known property for a star, then we could physically ``rescale'' the 
spectral energy distribution (SED), and infer, from the observed
flux, the distance of the body, $d$, or its absolute size ($R$), through 
a measure of the apparent angular extension, 
$\theta = (R/d)^2 \propto L\,T_{\rm eff}^{-4}\,d^{-2}$
\citep[][]{ridgway80,dyck96,perrin98,richichi98}.

As well known, however, $L$ cannot, in principle, be {\it directly} 
measured, requiring for this task an ideal detector equally sensitive to 
the whole spectral range.
The lack of this crucial piece of information is often palliated by
indirect observing methods, trying to pick up the  bulk of
stellar emission through broad-band photometry within the appropriate spectral
range according to target temperature.\footnote{Recalling that emission peak
roughly obeys the Wien law, i.e.\ $T \lambda_{\rm peak} \simeq {\rm const}$.}
Relying on this approach, \citet{johnson66} derived the bolometric vs.\ temperature scale
for red giant stars, while \citet{code76} explored the same
relation for hot early-type stars, through satellite-borne UV observations.
As an alternative way, many authors tried a fully theoretical assessment of
the problem, by studying the $f_{\rm bol}$ vs.\ $f_{\rm \lambda}$ 
relationship on the basis of model grids of stellar atmospheres and
replacing observations with synthetic photometry directly computed on the
theoretical SED \citep[][]{bertone04,bessell98}. 

Rather than focussing on luminosity, \citet{wesselink69} originally proposed 
a further application of this method, just looking at the bolometric 
surface brightness, namely $\mu = f_{\rm bol}/\theta^2$, to lead to a refined 
temperature scale of stars in force of the fundamental relationship 
$\mu = \sigma\,T_{\rm eff}^{-4}$ ($\sigma$ being the Stefan-Boltzmann constant). 
The so-called surface-brightness technique, then better 
recognized as the IR-flux method (IRFM), has been extensively applied 
to the study of red giant and supergiant stars \citep[][]{black79,dibenedetto87,black94,alonso99,ramirez05,gonzalez09}
taking advantage of its distance-independent results, providing to
match the angular measure of stellar radii with the estimate of the 
bolometric flux from infrared observations, i.e.\ $\mu = (f_{\rm bol}/f_{\rm IR}) f_{\rm IR}$.

Although in different forms, all the previous methods used theoretical models
of stellar atmospheres to derive the appropriate ``correcting factor''
${\cal R} = f_{\rm bol}/f(\lambda)$ and convert observed or synthetic 
monochromatic magnitudes, $m(\lambda)$ to the bolometric scale.\footnote{To a more
detailed analysis, note that the ratio ${\cal R}$ dimensionally matches
the definition of ``equivalent width'', and it gives
a measure of how ``broad'' is the whole SED compared to the monochromatic
emission density at the reference $\lambda$.}

\noindent Taking the Sun as a reference source for our calibration, 
we could write more explicitely: 
\begin{equation}
[m_{\rm bol} - m(\lambda)]- [m_{\rm bol} - m(\lambda)]_\odot 
= -2.5\,\log ({\cal R/R}_\odot)
\label{eq:bc}
\end{equation}

Equation (\ref{eq:bc}) actually leads to the straight definition 
of {\it bolometric correction}, $BC(\lambda)$, namely 
\begin{equation}
BC(\lambda) = (m_{\rm bol} - m(\lambda)) = -2.5\,\log {\cal R}
+ BC(\lambda)_\odot
\label{eq:bc2}
\end{equation}

Aside from the historical definition, that originally considered 
BC only to photographic ($m_{\rm pg}$) or visual 
($m_V$) magnitudes \citep{kuiper38}, one can nowadays easily extend the definition 
to any waveband. A careful analysis of eq.~(\ref{eq:bc2}) makes clear
some important properties of $BC$:
{\it i)} the value of $\cal R$ is a composite function of stellar
fundamental parameters, namely ${\cal R} = {\cal R}(T_{\rm eff},\log g, [X/H])$
so that, for fixed effective temperature, $BC$ may display some 
dependence on stellar gravity ($g$) and chemical composition ($[X/H]$);
{\it ii)} the value of $\cal R$ (and, accordingly, of $BC$) is minimum
when our observations catch the bulk of stellar luminosity.
For this reason, high values of $BC_V$ must be expected when observing
for instance cool giant stars in the $V$ band, or hot O-B stars in the
infrared $K$ band.
{\it iii)} The definition of the $BC$ scale strictly  depends on the assumed
reference value for the Sun, that therefore must univocally fix the ``zero point''
of the scale \citep{bessell98}.

In this framework, we want to tackle here the central question of the
possible $BC$ dependence on stellar metallicity. This effect could be 
of special importance, in fact, in order to more confidently set the 
bolometric vs.\ temperature scale
for cool red giants, where the intervening absorption of diatomic 
(TiO {\it in primis}) and triatomic (H$_2$O) molecules heavily
modulate the stellar SED with sizeable effects on optical and NIR 
magnitudes \citep[e.g.][]{gratton82,bertone08}. 
As a matter of fact, still nowadays the many efforts devoted to
the definition of the $BC$ vs.\ $\log T_{\rm eff}$ relationship 
led to non univocal conclusions, with large discrepancies among the 
different sources in the literature as far as stars of K spectral 
type or later are concerned \citep{flower75,flower77,bessell84,houd00,bertone04,worthey06}. 

This issue has actually an even more important impact
on the study of the integrated spectrophotometric properties
of resolved and unresolved stellar systems, as red giants and other 
Post-main sequence (PMS) stars provide a prevailing 
fraction \citep[2/3 or more,][]{buzzoni89}
of the total luminosity of the population. 
A fair definition of the $BC$ scale becomes therefore of paramount 
importance to self-consistently convert theoretical H-R and observed 
c-m diagrams of a stellar population \citep{flower96,vdb03} 
and more confidently assess the physical contribution of the different 
stellar classes.

A study of the $BC$ dependence on metal abundance has been previously 
attempted by many authors mainly relying on a fully theoretical point
of view to exploit the obvious advantage of stellar models
to account in a controlled way for a global or selective change of 
metal abundance. In this regard, \citet{tripicco95} and 
\citet{cassisi04}, among others, 
tried to explore the effect of $\alpha$ elements enhancement 
(namely O, Mg, Ca, Ti etc.) in stellar SED, while \citet{girardi07}
focussed on the possible impact of Helium abundance on $BC$.
As a major drawback of these efforts, however, one has to report the
admitted limit of model atmospheres in accurately describe the
spectrophotometric properties of K- and M-type stars, that are 
cooler than 4000~K \citep[see][on this important point]{bertone08}.

On the other hand, a fully empirical approach has been devised by
\citet{montegriffo98} and \citet{alonso99}, among others,
trying to reconstruct stellar SED, and therefrom infer the bolometric
flux, $f_{\rm bol}$, through optical broad-band photometry of stars 
in the Galactic field or in globular clusters. A recognized limit 
of these studies is, however, that they may suffer from the lack of coverage
of the stellar parameter space offered by the observations.
Moreover, as far as the cool-star sequence is concerned, optical 
multicolor photometry, alone, partially misses the bulk of stellar emission
(more centered toward the NIR spectral window); in addition, by
converting broad-band magnitudes into monochromatic flux densities, the
stellar SED is reconstructed at very poor spectral resolution, thus
possibly loosing important features that may bias the inferred bolometric
energy budget.

On this line, however, we want to further improve the analysis
proposing here more complete  spectroscopic observations 
for a large grid of red-giant stars in several Galactic clusters
along the entire metallicity scale from very metal-poor (i.e.\ 
$[Fe/H] \simeq -2.2$~dex) to super-solar ($[Fe/H] \simeq +0.4$~dex)
stellar populations. Our observations span the whole optical and 
NIR wavelength range, thus allowing a quite accurate shaping of
stellar SED. As we will demonstrate in the following of our discussion,
our procedure allowed us to sample about 70-90 \% of the total 
emission of our sample stars, thus leading to a virtually {\it direct}
measure of $f_{\rm bol}$, even for M-type stars 
as cool as 3500~K. 

We will arrange our discussion by presenting, in Sec.~2, our stellar database
together with further available information in the literature.
The analysis of the observing material will be assessed in more detail in Sec.~3, while
in Sec.~4 we will derive the SED for the whole sample leading to an estimate
of the effective temperature and bolometric correction for each star. 
The discussion of the inferred BC-color-temperature scale will be the focus of
Sec.~5, especially addressing the possible dependence of BC on stellar metallicity.
The comparison of our results with other relevant BC calibration in the
literature will also be carried out in this section, while in Sec.~6 we will summarize
the main conclusions of our work.

\section{Cluster database selection}

As we mainly aim at probing the impact of metallicity on the
BC of stars at the low-temperature regime,
a demanding constraint to set up our target sample was to 
explore a range as wide as possible in [Fe/H], and pick up
red giant stars with accurate measurements of their metallicity.
The cluster population in the Galaxy naturally provided the
ideal environment for our task. By combining
globular and open clusters one can easily span the whole
metallicity range pertinent to Pop I and II stars in our
and in external galaxies.
We therefore selected five template systems, namely
the three metal-poor globular clusters M15, M2 and M71, and two
metal-rich open clusters NGC~188 and NGC~6791 such as to
let metallicity span almost three orders of magnitude,
from [Fe/H]~$= -2.3$ up to +0.4.

For each cluster, a subset of $\sim 20$ suitable targets have
then been identified among the brightest and coolest red giants 
from the 2MASS infrared c-m diagram \citep{2mass}. 
In assembling the dataset we also took care of picking up those
objects out of more severely crowded regions of the clusters, and clearly
recognizable in bright asterisms such as to reduce the
chance of misidentification at the telescope.

\begin{figure}
\centerline{
\psfig{file=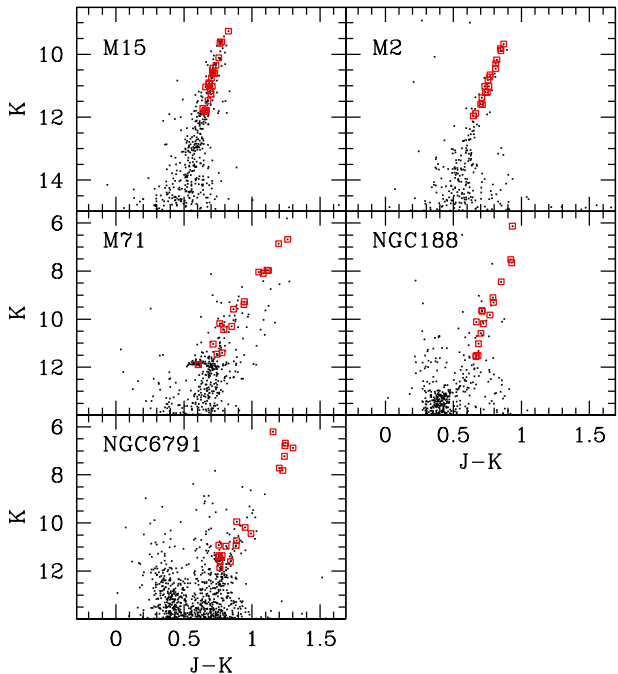,width=\hsize}
}
\caption{Apparent c-m diagram of the five clusters included in our analysis,
according to 2MASS J and K photometry. Big squares along the red giant branch
mark the selected targets in our sample.
}
\label{f0}
\end{figure}

The final set of target stars is summarized, for each cluster, 
in the five panels of Fig.~\ref{f0} and in the series of 
Tables \ref{t2} to \ref{t6}. We eventually 
considered 92 stars in total, of which 21 are in M15, 18 in M2, 17 
in M71, 16 in NGC~188, and 20 in NGC~6791, respectively. For each star, 
the tables always report the 2MASS id number (col.~1) and the alternative 
cross-identification, according to other reference photometric catalogs, 
when available.
The 2MASS J2000 coordinates on the sky and the corresponding $J,H,K$
magnitudes are also always reported, together with a compilation of 
$B,V,R_c,I_c$ observed magnitudes according to the best reference 
catalogs for each cluster, as reported in the literature.
When required, dereddened apparent magnitudes have been computed
according to the color excess $E(B-V)$ as labelled in the header
of each table.

\section{Observations and data reduction}

Spectroscopic observations of our stellar sample have been collected
during several runs between June and October 2003 at the 3.5m
Telecopio Nazionale Galileo (TNG) of the Roque de los Muchachos
Observatory, at La Palma (Canary Islands, Spain). A summary of the
logbook can be found in Table~\ref{t1}.

Optical spectroscopy was carried out with the LRS FOSC camera;
a composite spectrum was collected for each target by matching a 
blue (grism LRB along the $\lambda\lambda 3500-8800$~\AA\
wavelength range)\footnote{Although nominally extended to 9500~\AA, 
LRB spectra result severely affected by second-order spectral emission in
their red tail. For this reason, during data reduction, spectra 
have been clipped retaining only the wavelength region blueward 
of 8800~\AA.} and a red setup (grism LRR, between 
$\lambda\lambda 4500-10300$~\AA). In both cases the grisms
provided a dispersion of 2.8~\AA~px$^{-1}$ on a $2048\times 2048$ 
thinned and back-illuminated Loral CCD, with a 13.5$\mu$m pixel size.
In order to collect the entire flux from target stars,
we observed through a 5\arcsec wide slit; this condition actually
made spectral resolution to be eventually constrained
by the seeing figure (typically about 1-1.5\arcsec along the different
nights), thus ranging between 10 and 15~\AA\ (FWHM). This is equivalent to
a value of $R = \lambda/\Delta \lambda$ between 600-1000.
Whenever possible, and avoiding severe crowding conditions of the
target fields, the longslit was located at the parallactic angle.
Wavelength calibration and data reductions were performed following 
standard procedures.

\begin{table*}
\begin{minipage}{1.1\hsize}
\caption{Cluster properties and stellar database for cluster M~15}
\scriptsize{
\begin{tabular}{rrrllrrrrrrrr}
\hline
\multicolumn{13}{c}{\normalsize{\bf M~15:} $\qquad\quad$ \normalsize{\bf E(B--V) = 0.10} $\qquad\quad$ \normalsize{\bf [Fe/H] = --2.26}} \\
 &  &  &  &  &  &  &  &  &  &  &  \\
\multicolumn{3}{c}{ID} & \multicolumn{1}{c}{$\alpha$} & \multicolumn{1}{c}{$\delta$} &  
\multicolumn{1}{c}{B} & \multicolumn{2}{c}{\hrulefill ~~V~ \hrulefill} & \multicolumn{2}{c}{\hrulefill ~~I$_c$~ \hrulefill} &
\multicolumn{1}{c}{J} & \multicolumn{1}{c}{H} & \multicolumn{1}{c}{K} \\
\multicolumn{1}{c}{$^{(a)}$} & \multicolumn{1}{c}{$^{(b)}$} & \multicolumn{1}{c}{$^{(c)}$} & \multicolumn{2}{c}{(J2000.0)} &
\multicolumn{1}{c}{$^{(b)}$} & 
\multicolumn{1}{c}{$^{(b)}$} & \multicolumn{1}{c}{$^{(c)}$} & 
\multicolumn{1}{c}{$^{(b)}$} & \multicolumn{1}{c}{$^{(c)}$} & \multicolumn{1}{c}{$^{(a)}$} & 
\multicolumn{1}{c}{$^{(a)}$} & \multicolumn{1}{c}{$^{(a)}$} \\
\hline
21300002+1209182 & 165 &  71 & 21:30:00.02 & 12:09:18.24 & 15.334 & 14.395 & 14.3460 &13.330 & 13.2709 & 12.479 & 11.926 & 11.824 \\
21295705+1208531 & 959 &   6 & 21:29:57.06 & 12:08:53.11 & 14.549 & 13.426 & 13.4946 &12.144 & 12.2129 & 11.282 & 10.691 & 10.573 \\
21295532+1210327 & 337 &  60 & 21:29:55.33 & 12:10:32.80 & 15.229 & 14.313 & 14.3694 &13.165 & 13.2132 & 12.452 & 11.899 & 11.786 \\
21300090+1208571 & 558 & 461 & 21:30:00.91 & 12:08:57.13 & 14.092 & 12.700 & 12.9683 &11.281 & 11.4637 & 10.383 &  9.759 &  9.605 \\
21295473+1208592 & 330 &  25 & 21:29:54.73 & 12:08:59.24 & 14.821 & 13.691 & 13.7581 &12.444 & 12.5006 & 11.591 & 11.065 & 10.906 \\
21300461+1210327 & 369 &     & 21:30:04.62 & 12:10:32.73 & 14.851 & 13.836 &	     &       &         & 11.858 & 11.272 & 11.165 \\
21295560+1212422 & 533 & 665 & 21:29:55.61 & 12:12:42.29 & 14.562 & 13.459 & 13.5218 &       & 12.2237 & 11.336 & 10.723 & 10.609 \\
21300514+1210041 & 372 &     & 21:30:05.15 & 12:10:04.18 & 15.186 & 14.288 &	     &       &         & 12.430 & 11.929 & 11.776 \\
21295836+1209020 &     & 166 & 21:29:58.37 & 12:09:02.01 &	  &	   & 13.8205 &       & 12.5987 & 11.700 & 11.112 & 11.042 \\
21295618+1210179 &     & 631 & 21:29:56.18 & 12:10:17.93 &	  &	   & 12.7694 &       & 11.3768 & 10.414 &  9.781 &  9.649 \\
21295739+1209056 &     &   7 & 21:29:57.39 & 12:09:05.69 &	  &	   & 13.7397 &       & 12.5054 & 11.632 & 11.070 & 10.948 \\
21300097+1210375 &     &  65 & 21:30:00.98 & 12:10:37.60 &	  &	   & 13.8739 &       & 12.6289 & 11.726 & 11.171 & 11.017 \\
21300431+1210561 & 368 &     & 21:30:04.32 & 12:10:56.16 & 14.649 & 13.559 &	     &       &         & 11.459 & 10.893 & 10.757 \\
21301049+1210061 & 621 &     & 21:30:10.49 & 12:10:06.18 & 14.563 & 13.406 &	     &       &         & 11.151 & 10.562 & 10.438 \\
21300739+1210330 & 604 &     & 21:30:07.40 & 12:10:33.06 & 14.961 & 13.986 &	     &       &         & 11.964 & 11.399 & 11.264 \\
21300569+1210156 &     &     & 21:30:05.70 & 12:10:15.68 &	  &	   &	     &       &         & 12.156 & 11.596 & 11.480 \\
21300553+1208553 &     &     & 21:30:05.54 & 12:08:55.35 &	  &	   &	     &       &         & 12.357 & 11.835 & 11.719 \\
21295756+1209438 &     &     & 21:29:57.57 & 12:09:43.85 &	  &	   &	     &       &         & 10.096 &  9.429 &  9.269 \\
21295082+1211301 &     &     & 21:29:50.83 & 12:11:30.18 &	  &	   &	     &       &         & 11.326 & 10.725 & 10.612 \\
21295881+1209285 &     &  59 & 21:29:58.82 & 12:09:28.59 &	  &	   & 14.5465 &       & 13.5061 & 11.088 & 10.568 & 10.353 \\
21295716+1209175 &     & 273 & 21:29:57.17 & 12:09:17.52 &	  &	   & 13.1662 &       & 11.7880 & 10.867 & 10.220 & 10.112 \\
\hline
\end{tabular}

\smallskip
\small{$^{(a)}$ from 2MASS;\\ 
$^{(b)}$ from \citet{cohen05};\\ 
$^{(c)}$ from \citet{rosenberg00}\hfill}
}
\label{t2}
\end{minipage}
\end{table*}

The optical spectra have then been accompanied by the corresponding
observations taken at infrared wavelength with the NICS camera
at the Nasmyth focus of the TNG. The camera was coupled with a
Rockwell 1024$\times$1024 Hawaii-1 HgCdTe detector.
We took advantage of NICS unique design using the Amici grism
coupled with two slits 0.5\arcsec and 5\arcsec wide, 
the latter being used for a complete flux sampling
of the target stars. The spectra cover the entire wavelength 
range from 8000~\AA\ to 2.5~$\mu$m at a resolving power (for a 0.5\arcsec
slit) which varies between $R = 80$ and 140 along the spectrum.
In acquiring spectra, background subtraction and flat-fielding correction
were eased by a standard dithering procedure on target images,
while the wavelength calibration directly derived from the standard 
reference table providing the dispersion relation of the system.
The {\sc Midas} ESO package, and specifically its {\sc Longslit} routine
set, has been used for the whole reduction procedure, both for optical 
and infrared spectra.

\begin{table}
\begin{minipage}{1.1\hsize}
\caption{Cluster properties and stellar database for cluster M~2$^{(a)}$}
\scriptsize{
\begin{tabular}{rllrrr}
\hline
\multicolumn{6}{c}{\normalsize{\bf M~2:} $\qquad\quad$ \normalsize{\bf E(B--V) = 0.06} $\qquad\quad$ \normalsize{\bf [Fe/H] = --1.62}} \\
 &  &  &  &  &  \\
\multicolumn{1}{c}{ID} & \multicolumn{1}{c}{$\alpha$} & \multicolumn{1}{c}{$\delta$} &  
\multicolumn{1}{c}{J} & \multicolumn{1}{c}{H} & \multicolumn{1}{c}{K} \\
  & \multicolumn{2}{c}{(J2000.0)} &  &  &  \\
\hline
21333827-0054569 & 21:33:38.28 & --00:54:56.92 & 10.542 &  9.827 &  9.672 \\
21333095-0052154 & 21:33:30.96 & --00:52:15.47 & 11.568 & 10.952 & 10.814 \\
21332468-0044252 & 21:33:24.69 & --00:44:25.21 & 12.549 & 12.006 & 11.886 \\
21331771-0047273 & 21:33:17.71 & --00:47:27.31 & 10.665 &  9.961 &  9.821 \\
21331723-0048171 & 21:33:17.24 & --00:48:17.10 & 11.112 & 10.429 & 10.301 \\
21331790-0048198 & 21:33:17.91 & --00:48:19.82 & 11.746 & 11.103 & 11.017 \\
21331854-0051563 & 21:33:18.55 & --00:51:56.33 & 11.779 & 11.137 & 11.019 \\ 
21331948-0051034 & 21:33:19.49 & --00:51:03.42 & 11.963 & 11.299 & 11.214 \\
21331923-0049058 & 21:33:19.23 & --00:49:05.84 & 12.280 & 11.695 & 11.579 \\
21332588-0046004 & 21:33:25.89 & --00:46:00.44 & 12.313 & 11.756 & 11.600 \\
21333668-0051058 & 21:33:36.68 & --00:51:05.89 & 10.730 & 10.026 &  9.880 \\
21333520-0046089 & 21:33:35.21 & --00:46:08.91 & 10.993 & 10.324 & 10.174 \\
21333488-0047572 & 21:33:34.88 & --00:47:57.25 & 11.265 & 10.589 & 10.455 \\
21333593-0049224 & 21:33:35.94 & --00:49:22.44 & 11.420 & 10.750 & 10.650 \\
21333432-0051285 & 21:33:34.33 & --00:51:28.50 & 11.490 & 10.828 & 10.722 \\
21332531-0052511 & 21:33:25.32 & --00:52:51.17 & 11.938 & 11.300 & 11.203 \\
21333109-0054522 & 21:33:31.09 & --00:54:52.28 & 12.086 & 11.526 & 11.376 \\
21333507-0051097 & 21:33:35.07 & --00:51:09.72 & 12.609 & 12.056 & 11.962 \\
\hline
\end{tabular}

\smallskip
\small{$^{(a)}$ all the data are from 2MASS; \hfill}
}
\label{t3}
\end{minipage}
\end{table}

\begin{table*}
\begin{minipage}{1.1\hsize}
\caption{Cluster properties and stellar database for cluster M~71}
\scriptsize{
\begin{tabular}{rrrllrrrrrrr}
\hline
\multicolumn{12}{c}{\normalsize{\bf M~71:}$\qquad\quad$ \normalsize{\bf E(B--V) = 0.25}$\qquad\quad$ \normalsize{\bf [Fe/H] = --0.73}} \\
 &  &  &  &  &  &  &  &  &  &  &  \\
\multicolumn{3}{c}{ID} & \multicolumn{1}{c}{$\alpha$} & \multicolumn{1}{c}{$\delta$} &  
\multicolumn{1}{c}{B} & \multicolumn{2}{c}{\hrulefill ~~V~ \hrulefill} & \multicolumn{1}{c}{I$_c$} &
\multicolumn{1}{c}{J} & \multicolumn{1}{c}{H} & \multicolumn{1}{c}{K} \\
\multicolumn{1}{c}{$^{(a)}$} & \multicolumn{1}{c}{$^{(b)}$} & \multicolumn{1}{c}{$^{(c)}$} & \multicolumn{2}{c}{(J2000.0)} &
\multicolumn{1}{c}{$^{(b)}$} & 
\multicolumn{1}{c}{$^{(b)}$} & \multicolumn{1}{c}{$^{(c)}$} & 
\multicolumn{1}{c}{$^{(c)}$} & \multicolumn{1}{c}{$^{(a)}$} & 
\multicolumn{1}{c}{$^{(a)}$} & \multicolumn{1}{c}{$^{(a)}$} \\
\hline
19535325+1846471 & 2672 &  256 & 19:53:53.25 & 18:46:47.13 & 13.905 & 12.314 & 12.2085 & 10.3988 &  9.090 &  8.197 &  8.040 \\
19534750+1846169 & 2222 &  540 & 19:53:47.51 & 18:46:16.99 & 14.431 & 13.137 & 13.0010 & 11.5156 & 10.452 &  9.698 &  9.588 \\
19535150+1848059 & 2541 &  892 & 19:53:51.50 & 18:48:05.91 & 14.079 & 12.436 & 12.3250 & 10.4275 &  9.079 &  8.207 &  7.968 \\
19535064+1849075 & 2461 &  331 & 19:53:50.64 & 18:49:07.52 & 14.466 & 13.064 & 12.9955 & 11.4204 & 10.215 &  9.446 &  9.271 \\
19534575+1847547 & 2079 &  648 & 19:53:45.76 & 18:47:54.80 & 14.247 & 12.606 & 12.4924 & 10.5109 &  9.094 &  8.203 &  7.974 \\
19534827+1848021 & 2281 &  309 & 19:53:48.27 & 18:48:02.17 & 14.078 & 12.492 & 12.3636 & 10.5500 &  9.177 &  8.270 &  8.094 \\
19534656+1847441 & 2145 &   46 & 19:53:46.57 & 18:47:44.19 & 14.838 & 13.623 & 13.5524 & 12.2176 & 11.228 & 10.569 & 10.435 \\
19535369+1846039 & 2711 &  172 & 19:53:53.70 & 18:46:03.98 & 15.527 & 14.578 & 14.4974 & 13.3402 & 12.500 & 11.998 & 11.896 \\  				     
19534905+1846003 & 2337 &  303 & 19:53:49.05 & 18:46:00.34 & 14.601 & 13.410 & 13.3436 & 11.9991 & 10.950 & 10.276 & 10.186 \\
19534916+1846512 & 2347 &    6 & 19:53:49.16 & 18:46:51.22 & 14.997 & 13.709 & 13.6219 & 12.2031 & 11.151 & 10.434 & 10.301 \\
19534178+1848384 & 1772 &      & 19:53:41.79 & 18:48:38.46 & 15.877 & 14.694 &         &	 & 12.183 & 11.521 & 11.402 \\
19535676+1845399 & 2921 &      & 19:53:56.77 & 18:45:39.95 & 15.747 & 14.605 &         &	 & 12.197 & 11.529 & 11.455 \\
19533962+1848569 & 1611 &      & 19:53:39.62 & 18:48:56.99 & 15.695 & 14.627 &         &	 & 12.494 & 11.974 & 11.888 \\
19533864+1847554 & 1543 &      & 19:53:38.64 & 18:47:55.45 & 15.475 & 14.222 &         &	 & 11.751 & 11.151 & 11.037 \\
19534615+1847261 &	&  580 & 19:53:46.15 & 18:47:26.11 &	    &	     & 13.1140 & 11.5109 & 10.336 &  9.543 &  9.395 \\ 
19535610+1847167 & 2885 & 1066 & 19:53:56.10 & 18:47:16.76 & 13.577 & 11.905 & 12.4009 &  9.2167 &  7.943 &  7.078 &  6.681 \\
19534941+1844269 & 2365 &      & 19:53:49.41 & 18:44:26.98 & 13.863 & 12.107 &         &	 &  8.058 &  7.105 &  6.863 \\
\hline
\end{tabular}

\smallskip
\small{$^{(a)}$ from 2MASS;\\ 
$^{(b)}$ from \citet{geffert00};\\ 
$^{(c)}$ from \citet{rosenberg00}\hfill}
}
\label{t4}
\end{minipage}
\end{table*}

\begin{table*}
\begin{minipage}{1.1\hsize}
\caption{Cluster properties and stellar database for NGC~188}
\scriptsize{
\begin{tabular}{rrlllrrrrrrrrr}
\hline
\multicolumn{14}{c}{\normalsize{\bf NGC 188:} $\qquad\quad$ \normalsize{\bf E(B--V) = 0.082} $\qquad\quad$ \normalsize{\bf [Fe/H] = --0.02}} \\
 &  &  &  &  &  &  &  &  &  &  &  \\
\multicolumn{3}{c}{ID} & \multicolumn{1}{c}{$\alpha$} & \multicolumn{1}{c}{$\delta$} &  
\multicolumn{2}{c}{\hrulefill ~~B~ \hrulefill} & \multicolumn{2}{c}{\hrulefill ~~V~ \hrulefill} & \multicolumn{1}{c}{R$_c$} & \multicolumn{1}{c}{I$_c$} &
\multicolumn{1}{c}{J} & \multicolumn{1}{c}{H} & \multicolumn{1}{c}{K} \\
\multicolumn{1}{c}{$^{(a)}$} & \multicolumn{1}{c}{$^{(b)}$} & \multicolumn{1}{c}{$^{(c)}$} & \multicolumn{2}{c}{(J2000.0)} &
\multicolumn{1}{c}{$^{(b)}$} & \multicolumn{1}{c}{$^{(c)}$} & 
\multicolumn{1}{c}{$^{(b)}$} & \multicolumn{1}{c}{$^{(c)}$} & 
\multicolumn{1}{c}{$^{(c)}$} & \multicolumn{1}{c}{$^{(c)}$} & \multicolumn{1}{c}{$^{(a)}$} & 
\multicolumn{1}{c}{$^{(a)}$} & \multicolumn{1}{c}{$^{(a)}$} \\
\hline
00445253+8514055 & 4668 & N188-I-69   &00:44:52.54 & 85:14:05.54  & 13.613 & 13.579 &  12.319 & 12.357 & 11.598 & 11.087 & 10.098 &  9.461 &  9.304 \\
00475922+8511322 & 5887 & N188-II-181 &00:47:59.23 & 85:11:32.28  & 13.587 & 13.428 &  12.135 & 12.197 & 11.429 & 10.894 &  9.891 &  9.203 &  9.100 \\
00465966+8513157 & 5085 & N188-I-105  &00:46:59.66 & 85:13:15.71  & 13.603 & 13.538 &  12.362 & 12.422 & 11.732 & 11.269 & 10.349 &  9.789 &  9.639 \\
00453697+8515084 & 5927 & N188-I-57   &00:45:36.97 & 85:15:08.43  & 14.799 & 14.760 &  13.658 & 13.706 & 13.039 & 12.571 & 11.709 & 11.149 & 11.024 \\
00442946+8515093 & 4636 & N188-I-59   &00:44:29.46 & 85:15:09.39  & 14.986 & 14.950 &  14.005 & 14.046 & 13.385 & 12.962 & 12.202 & 11.653 & 11.520 \\
00473222+8511024 & 5133 & N188-II-187 &00:47:32.22 & 85:11:02.45  & 15.171 & 15.132 &  14.077 & 14.140 & 13.490 & ...	 & 12.234 & 11.700 & 11.567 \\
00554526+8512209 & 6175 &	      &00:55:45.27 & 85:12:20.92  & 12.224 & ...    &  10.834 & ...    & ...	& ...	 &  8.441 &  7.631 &  7.520 \\
00463920+8523336 & 4843 &	      &00:46:39.21 & 85:23:33.67  & 12.890 & ...    &  11.569 & ...    & ...	& ...	 &  9.292 &  8.597 &  8.441 \\
00472975+8524140 & 4829 &	      &00:47:29.76 & 85:24:14.09  & 13.965 & ...    &  12.781 & ...    & ...	& ...	 & 10.783 & 10.210 & 10.114 \\
00441241+8509312 & 4756 &	      &00:44:12.42 & 85:09:31.23  & 12.933 & ...    &  11.404 & ...    & ...	& ...	 &  8.580 &  7.892 &  7.652 \\
00432696+8509175 & 4408 &	      &00:43:26.96 & 85:09:17.58  & 14.242 & ...    &  13.199 & ...    & ...	& ...	 & 11.293 & 10.706 & 10.591 \\
00471847+8519456 & 4909 &	      &00:47:18.48 & 85:19:45.65  & 14.255 & ...    &  13.010 & ...    & ...	& ...	 & 10.908 & 10.289 & 10.187 \\
00461981+8520086 & 4524 &	      &00:46:19.81 & 85:20:08.61  & 13.663 & ...    &  12.468 & ...    & ...	& ...	 & 10.385 &  9.816 &  9.674 \\
00463004+8511518 & 5894 &	      &00:46:30.05 & 85:11:51.89  & 15.142 & ...    &  14.052 & ...    & ...	& ...	 & 12.185 & 11.695 & 11.518 \\
00490560+8526077 & 5835 &	      &00:49:05.60 & 85:26:07.77  & 13.921 & ...    &  12.717 & ...    & ...	& ...	 & 10.594 &  9.956 &  9.825 \\
00420323+8520492 & \multicolumn{2}{l}{$\equiv$ SAO~109} &00:42:03.23 & 85:20:49.23  & \multicolumn{2}{c}{11.40$^{(d)}$}  & \multicolumn{2}{c}{9.89$^{(d)}$} & ...  & ...    &  7.064 &  6.387 &  6.130 \\
\hline
\end{tabular}

\smallskip
\small{$^{(a)}$ from 2MASS;\\ 
$^{(b)}$ from \citet{platais03};\\ 
$^{(c)}$ from \citet{stetson04};\\
$^{(d)}$ B and V photometry from SIMBAD\hfill}
}
\label{t5}
\end{minipage}
\end{table*}

\begin{table*}
\begin{minipage}{1.1\hsize}
\caption{Cluster properties and stellar database for cluster NGC~6791}
\scriptsize{
\begin{tabular}{rrrllrrrrrrrr}
\hline
\multicolumn{13}{c}{\normalsize{\bf NGC 6791:} $\qquad\quad$ \normalsize{\bf E(B--V) = 0.117} $\qquad\quad$ \normalsize{\bf [Fe/H] = +0.4}} \\
 &  &  &  &  &  &  &  &  &  &  &  \\
\multicolumn{3}{c}{ID} & \multicolumn{1}{c}{$\alpha$} & \multicolumn{1}{c}{$\delta$} &  
\multicolumn{2}{c}{\hrulefill ~~B~ \hrulefill} & \multicolumn{2}{c}{\hrulefill ~~V~ \hrulefill} & \multicolumn{1}{c}{I$_c$} &
\multicolumn{1}{c}{J} & \multicolumn{1}{c}{H} & \multicolumn{1}{c}{K} \\
\multicolumn{1}{c}{$^{(a)}$} & \multicolumn{1}{c}{$^{(b)}$} & \multicolumn{1}{c}{$^{(c)}$} & \multicolumn{2}{c}{(J2000.0)} &
\multicolumn{1}{c}{$^{(b)}$} & \multicolumn{1}{c}{$^{(c)}$} & 
\multicolumn{1}{c}{$^{(b)}$} & \multicolumn{1}{c}{$^{(c)}$} & 
\multicolumn{1}{c}{$^{(c)}$} & \multicolumn{1}{c}{$^{(a)}$} & 
\multicolumn{1}{c}{$^{(a)}$} & \multicolumn{1}{c}{$^{(a)}$} \\
\hline
19210807+3747494 &  6697 & 3475 & 19:21:08.07 & 37:47:49.41 &  15.279 & 15.275 & 13.909 & 13.978 & 12.668 & 11.675 & 11.071 & 10.919 \\
19204971+3743426 & 10807 & 3502 & 19:20:49.72 & 37:43:42.67 &  15.554 & 15.563 & 13.956 & 13.982 & 10.990 &  9.041 &  8.167 &  7.815 \\
19205259+3744281 & 10140 & 2228 & 19:20:52.60 & 37:44:28.18 &  15.715 & 15.732 & 14.095 & 14.150 & 12.397 & 11.135 & 10.417 & 10.185 \\
19205580+3742307 & 11799 & 3574 & 19:20:55.81 & 37:42:30.75 &  16.307 & 16.297 & 14.934 & 14.957 & 13.609 & 12.622 & 11.993 & 11.860 \\
19205671+3743074 & 11308 & 2478 & 19:20:56.72 & 37:43:07.46 &  16.000 & 15.984 & 14.633 & 14.660 & 13.325 & 12.351 & 11.756 & 11.586 \\
19210112+3742134 & 12010 & 3407 & 19:21:01.12 & 37:42:13.45 &  15.942 & 15.928 & 14.433 & 14.455 & 12.901 & 11.821 & 11.130 & 10.938 \\
19211606+3746462 &  7750 &	& 19:21:16.06 & 37:46:46.26 &  15.472 &        & 13.871 &	 &	  &  8.914 &  8.053 &  7.714 \\
19213656+3740376 & 12650 &	& 19:21:36.56 & 37:40:37.63 &  15.727 &        & 14.174 &	 &	  & 11.431 & 10.635 & 10.438 \\
19210326+3741190 & 13637 &	& 19:21:03.27 & 37:41:19.04 &  15.722 &        & 14.348 &	 &	  & 12.120 & 11.516 & 11.362 \\
19213635+3739445 & 13082 &	& 19:21:36.36 & 37:39:44.57 &  16.186 &        & 14.825 &	 &	  & 12.449 & 11.728 & 11.608 \\
19212437+3735402 & 15790 &	& 19:21:24.37 & 37:35:40.29 &  15.837 &        & 14.442 &	 &	  & 12.134 & 11.546 & 11.354 \\
19212674+3735186 &	 &	& 19:21:26.75 & 37:35:18.60 &	      &        &	&	 &	  & 11.622 & 10.925 & 10.735 \\
19211632+3752154 &  3254 &	& 19:21:16.32 & 37:52:15.46 &  15.282 &        & 13.998 &	 &	  & 11.776 & 11.068 & 10.967 \\
19211176+3752459 &  2970 &	& 19:21:11.76 & 37:52:46.00 &  15.676 &        & 14.336 &	 &	  & 12.174 & 11.557 & 11.420 \\
19202345+3754578 &  1829 &	& 19:20:23.45 & 37:54:57.82 &  14.592 &        & 12.866 &	 &	  &  8.029 &  7.133 &  6.787 \\
19205149+3739334 &	 &	& 19:20:51.50 & 37:39:33.44 &	      &        &	&	 &	  &  7.356 &  6.516 &  6.201 \\
19203285+3753488 &  2394 &	& 19:20:32.85 & 37:53:48.87 &  15.056 &        & 13.417 &	 &	  &  8.463 &  7.535 &  7.224 \\
19200641+3744452 &  9800 &	& 19:20:06.42 & 37:44:45.28 &  14.670 &        & 13.307 &	 &	  & 10.831 & 10.094 &  9.943 \\
19200882+3744317 & 10034 &	& 19:20:08.83 & 37:44:31.71 &  15.353 &        & 13.710 &	 &	  &  7.916 &  6.989 &  6.670 \\
19203219+3744208 & 10223 &	& 19:20:32.20 & 37:44:20.81 &  16.421 &        & 14.854 &	 &	  &  8.176 &  7.262 &  6.874 \\
\hline
\end{tabular}

\smallskip
\small{$^{(a)}$ from 2MASS;\\ 
$^{(b)}$ from \citet{kaluzny95};\\ 
$^{(c)}$ from \citet{stetson03}\hfill}
}
\label{t6}
\end{minipage}
\end{table*}

\subsection{ Flux calibration }

Given the nature of our investigation, special care has been devoted
to suitably fluxing both optical and infrared spectra. This has
been carried out by repeated observations, both with LRS and NICS,
of a grid of spectrophotometric standard stars from the list of
\citet{massey88} and \citet{hunt98}, as reported in Table~\ref{t1}.
Note, however, that the lack of an appropriate SED calibration
of standard stars along the entire wavelength range of our
observations required a two-step procedure, relying on the direct
observation of Vega as a primary calibrator, according to
\citet{tokunaga05} results. Given the outstanding luminosity
of this star we had to observe through a 10 mag neutral 
filter to avoid CCD saturation, and create a secondary calibrator 
(namely HD192281) observed both with and without the neutral density
filter.

Concerning the applied correction for atmosphere absorption,
we had to manage two delicate problems. From one hand, in fact,
the intervening action of Sahara dust (the so-called ``calima effect'')
may abruptly increase the atmosphere opacity at optical wavelength.
This is a recurrent feature for summer nights at La Palma, and
it can severily affect the observing output, especially when
dealing with absolute flux calibration. A careful check with repeated
observations of the same standard stars along each night
allowed us to assess the presence of dust in the air. This confirmed,
for instance, that along our observing runs, the night of Aug 07, 2003
displayed an outstanding (i.e.\ a factor of four higher than the average)
dust extinction.

On the other hand, atmosphere water vapour can also play a
role by affecting in unpredictable ways the infrared observations.
Telluric H$_2$O bands about 1.10, 1.38 and 1.88~$\mu$m
\citep{fuensalida98,manduca79}, just restraining
to the Amici wavelength range, may in fact strongly contaminate the
{\it intrinsic} H$_2$O absorption bands of stellar SED, especially
for stars cooler than 3500~K \citep{bertone08}.
This effect may act on short timescales along the night, so that
it cannot be reconducted to an average nightly extinction curve,
as for optical observations.
The H$_2$O contamination in each spectrum was therefore corrected by 
re-scaling the average extinction curve to minimize the residual water
vapour feature in the stellar spectra.

\begin{table}
\caption{Logbook of TNG observations along 2003}
\scriptsize{
\begin{tabular}{llll}
\hline
Obs. date & Instrument & Targets & Standards$^\dagger$\\
 (2003)   &            &         &          \\
\hline
Jul 29 & LRS  & NGC6791 & HD192281\\
Jul 30 & LRS  & NGC6791 & HD192281, SAO48300, WOLF1346\\
Jul 31 & LRS  & NGC6791 & HD192281, SAO48300, WOLF1346\\
Aug 6  & LRS  & M71     & HD192281, SAO48300, WOLF1346\\
Aug 7  & LRS  & M15     & HD192281, SAO48300, WOLF1346\\
Aug 11 & NICS &         & HD192281, SAO48300\\
Aug 12 & NICS &         & Vega\\
Aug 18 & NICS & M71     & HD192281\\
Aug 19 & NICS &         & Vega\\
Aug 20 & NICS & M15, M71,& HD192281, SAO48300, WOLF1346,\\
       &      &  NGC188 &  Vega \\
Aug 21 & LRS  & M2      & HD192281\\
Aug 23 & LRS  & M2, M15 & HD192281\\
Aug 26 & LRS  & NGC188  &         \\
Aug 27 & LRS  & NGC188  & HD192281\\
Aug 31 & NICS & M71     & HD192281\\
Sep 1  & NICS & M71, NGC6791 & HD192281\\
Sep 3  & NICS & M2, M15, NGC188 & SAO48300\\
Sep 4  & NICS & NGC188 & HD192281\\
Sep 5  & NICS & NGC188, NGC6791 & HD192281\\
Oct 14 & NICS & M15   & HD192281\\
Oct 15 & NICS & M2    & HD192281\\
\hline
\end{tabular}

\smallskip
\small{$^\dagger$ HD192281 and WOLF1346 from optical calibration by 
\citet{massey88}; SAO48300 from JHK photometric calibration by 
\citet{hunt98}; Vega from \citet{tokunaga05}\hfill}
}
\label{t1}
\end{table}

Allover, the full calibration procedure led us to consistently
assemble the LRB-LRR-Amici spectral branches and obtain a nominal
SED of target stars along the 3450-25000~\AA\ wavelength range.
However, just an eye inspection of the full spectra made evident
in some cases a residual systematic component causing a
``glitch'' at the boundary connection between LRS and NICS observations.
Clearly, this effect urged us to further refine our analysis taking into
account the supplementary photometric piece of information,
as we will discuss in more detail in the next section.

\begin{figure}
\psfig{file=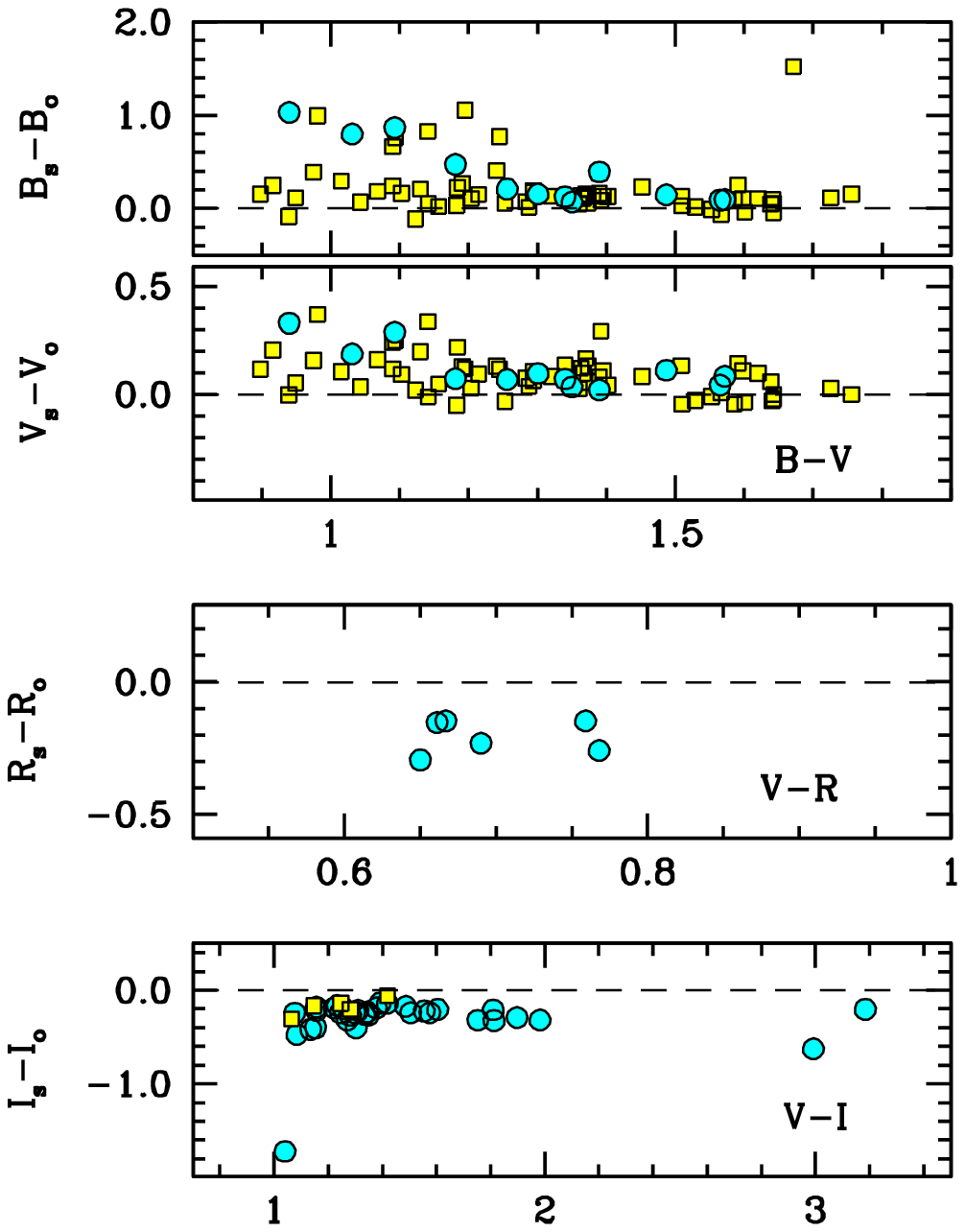,width=\hsize}
\psfig{file=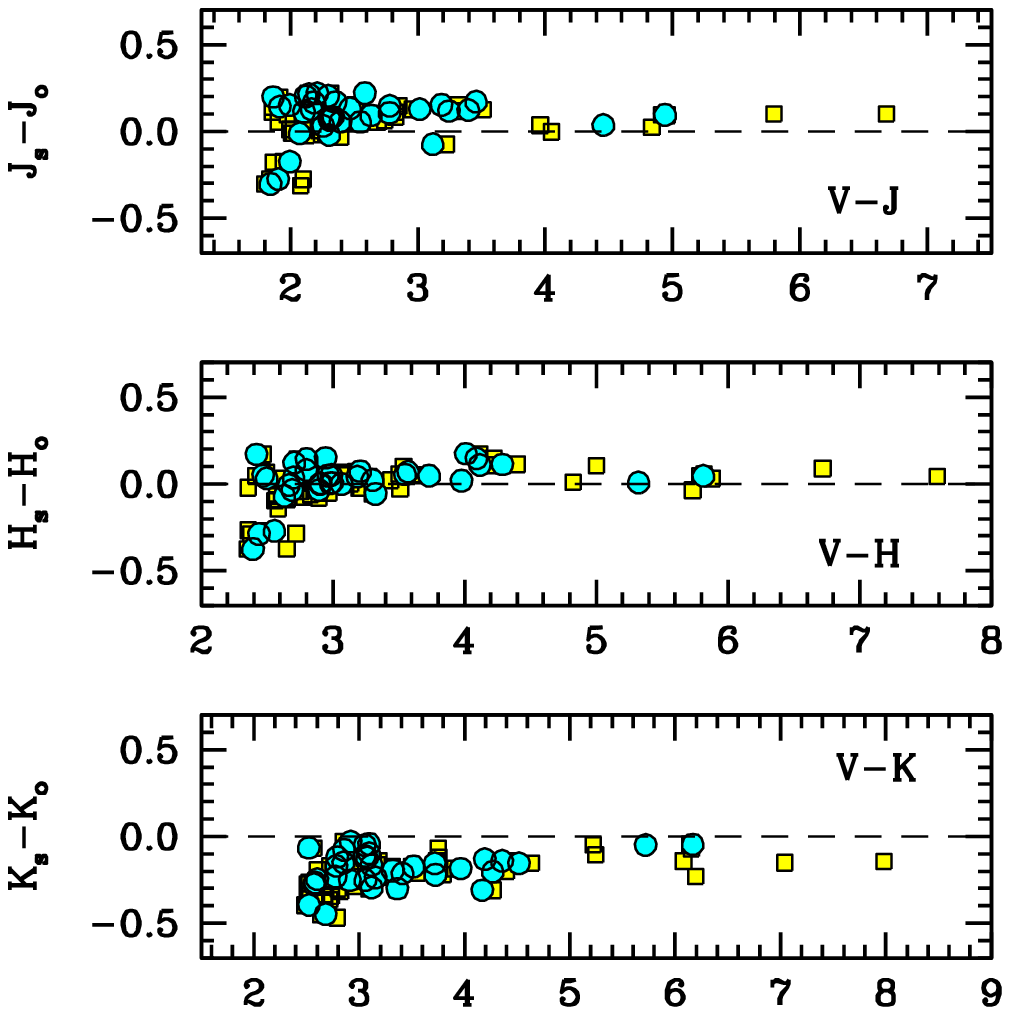,width=\hsize}
\caption{The $B,V,R_c,I_c,J,H,K$ magnitude residuals between synthetic 
and observed magnitudes (in the sense ``syn'' - ``obs'') for the 94 stars 
in our sample.plotted vs.\ literature colors, according to the data of
Table~\ref{t2} to \ref{t6}.
Synthetic magnitudes derived from the numerical integration of the observed 
SED with the Johnson-Cousins filters. All the available photometry has been accounted 
for. Some stars with multiple $V$ datasets appear therefore twice in the 
plots and are singled out by dot and square markers, respectively.
}
\label{f1}
\end{figure}

\subsection{Photometry and spectral ``fine tuning''}

The relevant database of broad-band photometry available in the
literature for all stars in our sample can be usefully accounted
for our analysis as a supplementary tool to tackle the inherent difficulty
in reproducing the overall shape of stellar SED at the required accuracy
level over the entire range of our observations.

As summarized in Tables \ref{t2} to \ref{t6}, a wide collection of photometric
catalogs can be considered, providing multicolor photometry along
the range spanned by LRS and NICS spectra. Facing the observed values, 
one can similarly derive a corresponding set of multicolor synthetic 
magnitudes relying on the assembled SED of each star. Operationally, 
from our $f(\lambda)$ values we need to  numerically assess the quantity

\begin{equation}
m_{\rm syn}^j = -2.5\,\log {{\int f(\lambda)S(\lambda)^j\,d\lambda}\over
{\int S(\lambda)^j\,d\lambda}} -2.5\,\log f_o^j
\label{eq:mag}
\end{equation}
being $m_{\rm syn}^j$ the synthetic magnitude in the {\it ``j-th''}
photometric band, identified by a filter response $S(\lambda)^j$ and
a calibrating zero-point flux $f_o^j$. For our calculations we relied on
the \citet{buzzoni05} reference data (see Table~1 therein).

A comparison of our output with the available photometry is displayed
in Fig.~\ref{f1}. The magnitude difference (in the sense 
``synthetic'' - ``observed''), is plotted in the different panels 
of the figure vs.\ {\it observed} color,
according to the different photometric catalogs quoted in Tables \ref{t2} 
to \ref{t6}.
As typically two sources for $V$ magnitudes are available for most
clusters, observed colors have been computed for each available $V$ dataset
and are displayed with a different marker (either dot or square) in the plots.

\begin{figure}
\psfig{file=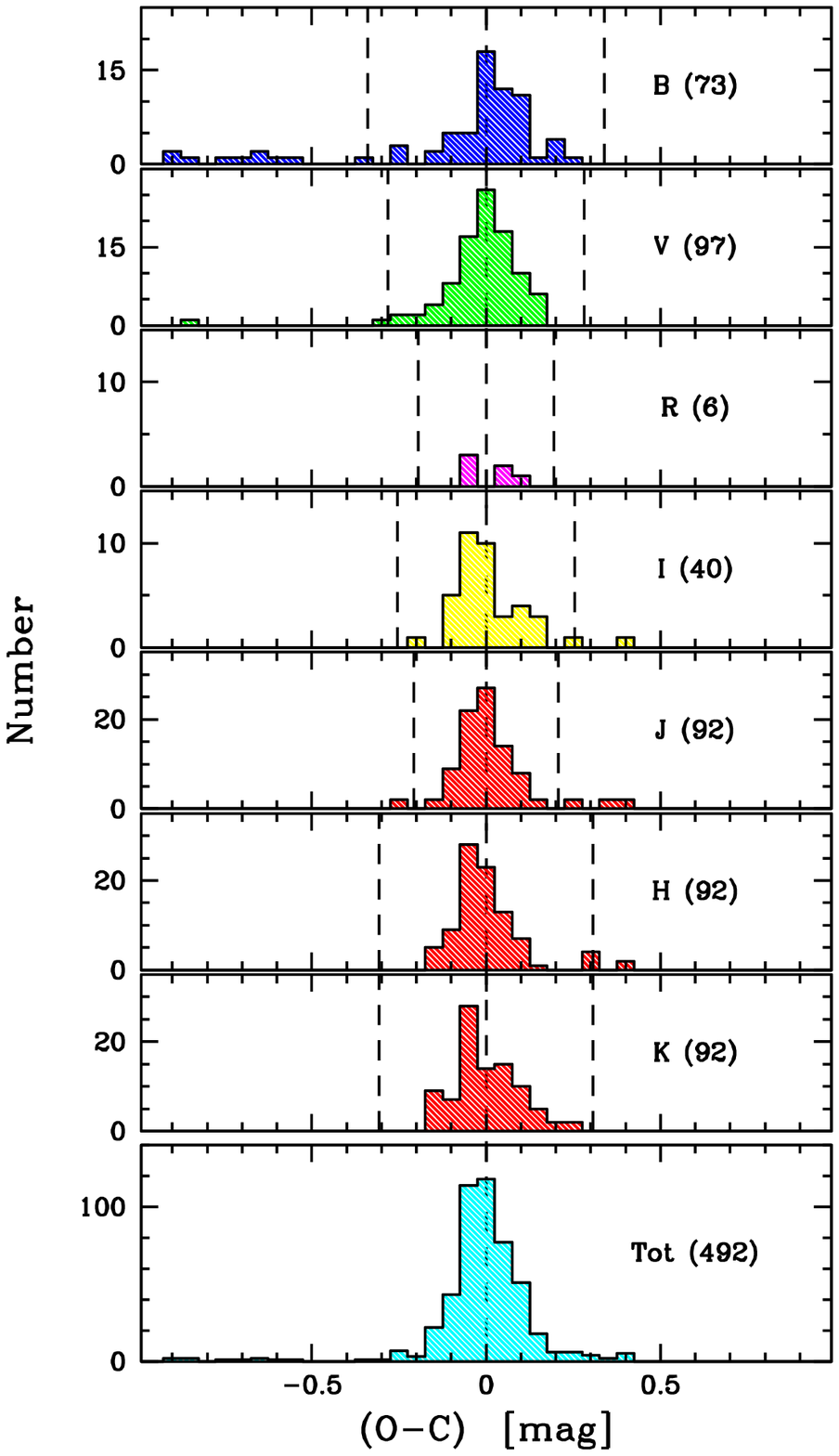,width=0.82\hsize}
\caption{The histogram of magnitude residuals for the data of Fig.~\ref{f1},
after correction for the systematic offsets, according to Table~\ref{t7}.
A total of 492 measures
have been accounted for, as labelled in the global histogram of the bottom panel,
including multiple photometry sources in the literature from Tables \ref{t2} to \ref{t6}.
Dashed vertical lines mark the $\pm 3\sigma$ clipping edges, according to our
iterative procedure, as devised in Sec.~3.2. Mag residuals are in the sense
``obs'' -- ``syn''. After outliers rejections, the global sample of 458 measurements
has, on average, $\sigma(\Delta{\rm mag}) = \pm 0.095$ (see Table~\ref{t7}).
}
\label{f2}
\end{figure}

Just a glance to Fig.~\ref{f1} makes evident that systematic offsets are
present between observed photometry and synthetic magnitudes.
This may partly be due to zero-point uncertainty in computing 
eq.~(\ref{eq:mag}), as well as to residual systematic drifts inherent 
to our spectral flux calibration. In addition, from the figure  one 
has also to report a few outliers in every band, and a notably skewed 
distribution of $B$ residuals.
To recover for this systematics we devised an iterative $3\sigma$
clipping procedure on the data of Fig.~\ref{f1} to reject deviant stars 
and lead synthetic magnitudes to match the standard photometric system
of the observed catalogs. Our results are displayed in graphical 
form in the plots of Fig.~\ref{f2}.

After just a few rejections, our procedure quickly converged to mean 
magnitude offsets
($\langle {\rm Obs} - {\rm Syn} \rangle$, see Table~\ref{t7}) to correct 
eq.~(\ref{eq:mag}) output. After correction for this systematics, our final 
synthetic photometry of cluster stars (not accouting  
for Galactic reddening) is collected in Tables~\ref{t8} and \ref{t9}.
According to Table~\ref{t7}, note that a $\sigma = 0.095$~mag in total magnitude
residuals evidently implies an {\it internal} accuracy in our spectral 
flux calibration of target stars better than 10\%.

\begin{table}
\caption{Magnitude residuals between observed and theoretical magnitudes$^{(a)}$}
\scriptsize{
\begin{tabular}{lcrrr}
\hline 
  & \multicolumn{4}{c}{All the clusters} \\
\multicolumn{1}{c}{Band} & \multicolumn{1}{c}{$\langle Obs - Syn \rangle$} & \multicolumn{1}{c}{$\sigma$} & N$_{\rm in}$ & N$_{\rm out}$ \\
\hline
B      & -0.137 &  0.113 &  63 &  10 \\
V      & -0.091 &  0.094 &  72 &   1 \\
R$_c$  & +0.205 &  0.065 &   6 &   - \\
I$_c$  & +0.244 &  0.085 &  38 &   2 \\
J      & -0.093 &  0.069 &  88 &   9 \\
H      & -0.012 &  0.102 &  94 &   3 \\
K      & +0.197 &  0.102 &  97 &   - \\
\hline
total  &  0.000$^{(b)}$ &  0.095 & 458 &  25 \\
\hline
\end{tabular}

\smallskip
\small{(a) Mag residuals are in the sense of observed -- synthetic one\\
(b) Weigthing with the number of entries, $N_{\rm in}$. \hfill}
}
\label{t7}
\end{table}

\begin{table}
\caption{Standard synthetic photometry from SED of target stars in globular clusters M~71, M~15, and M~2$^{(a)}$}
\scriptsize{
\begin{tabular}{lrrrrrrr}
\hline
\multicolumn{8}{c}{\large{\bf M~15}}\\
\multicolumn{1}{c}{ID} & \multicolumn{1}{c}{B} & \multicolumn{1}{c}{V} & \multicolumn{1}{c}{R$_c$} & \multicolumn{1}{c}{I$_c$} &
\multicolumn{1}{c}{J} & \multicolumn{1}{c}{H} & \multicolumn{1}{c}{K} \\
\hline
21300002+1209182 & 15.11  & 14.30  & 13.81  & 13.27  & 12.58  & 12.08  & 11.95 \\ 
21295705+1208531 & 14.30  & 13.35  & 12.79  & 12.18  & 11.41  & 10.82  & 10.74 \\ 
21295532+1210327 & 15.34  & 14.43  & 13.85  & 13.24  & 12.50  & 11.94  & 11.72 \\ 
21300090+1208571 & 14.04  & 12.90  & 12.21  & 11.47  & 10.51  &  9.82  &  9.50 \\ 
21295473+1208592 & 14.89  & 13.80  & 13.19  & 12.55  & 11.62  & 11.02  & 10.95 \\ 
21300461+1210327 & 15.01  & 13.85  & 13.24  & 12.62  & 11.86  & 11.16  & 10.98 \\ 
21295560+1212422 & 14.59  & 13.46  & 12.86  & 12.23  & 11.41  & 10.79  & 10.55 \\ 
21300514+1210041 & 15.20  & 14.31  & 13.79  & 13.23  & 12.45  & 11.89  & 11.64 \\ 
21295836+1209020 & 14.80  & 13.83  & 13.27  & 12.65  & 11.81  & 11.22  & 11.01 \\ 
21295618+1210179 & 14.02  & 12.89  & 12.19  & 11.48  & 10.49  &  9.82  &  9.55 \\ 
21295739+1209056 & 14.89  & 13.83  & 13.22  & 12.58  & 11.65  & 11.05  & 11.03 \\ 
21300097+1210375 & 14.91  & 13.85  & 13.24  & 12.63  & 11.85  & 11.19  & 11.14 \\ 
21300431+1210561 & 14.76  & 13.59  & 12.91  & 12.25  & 11.40  & 10.81  & 10.65 \\ 
21301049+1210061 & 14.45  & 13.36  & 12.73  & 12.09  & 11.25  & 10.60  & 10.35 \\ 
21300739+1210330 & 15.22  & 14.05  & 13.36  & 12.67  & 11.86  & 11.24  & 11.10 \\ 
21300569+1210156 & 15.02  & 14.11  & 13.56  & 12.97  & 12.21  & 11.61  & 11.50 \\ 
21300553+1208553 & 15.82  & 14.84  & 14.26  & 13.61  & 12.50  & 11.86  & 11.65 \\ 
21295756+1209438 & 13.80  & 12.45  & 11.72  & 11.01  & 10.10  &  9.47  &  9.32 \\ 
21295082+1211301 & 14.61  & 13.52  & 12.91  & 12.27  & 11.40  & 10.82  & 10.66 \\ 
21295881+1209285$^{(b)}$ & 14.62  & 13.40  & 12.70  & 12.03  & 11.17  & 10.57  & 10.42 \\ 
21295716+1209175 & 13.96  & 13.03  & 12.45  & 11.84  & 10.98  & 10.36  & 10.20 \\ 
\hline
\hline
\multicolumn{8}{c}{\large{\bf M~2}}\\
\multicolumn{1}{c}{ID} & \multicolumn{1}{c}{B} & \multicolumn{1}{c}{V} & \multicolumn{1}{c}{R$_c$} & \multicolumn{1}{c}{I$_c$} &
\multicolumn{1}{c}{J} & \multicolumn{1}{c}{H} & \multicolumn{1}{c}{K} \\
\hline
21333827-0054569 & 13.71  & 12.72  & 12.17  & 11.58  & 10.55  &  9.85  &  9.73 \\ 
21333095-0052154 & 14.74  & 13.68  & 13.11  & 12.53  & 11.59  & 10.96  & 10.87 \\ 
21332468-0044252 & 15.71  & 14.65  & 14.07  & 13.48  & 12.59  & 12.01  & 11.94 \\ 
21331771-0047273 & 14.30  & 13.02  & 12.32  & 11.66  & 10.69  &  9.94  &  9.91 \\ 
21331723-0048171 & 15.06  & 13.66  & 12.91  & 12.21  & 11.18  & 10.44  & 10.32 \\ 
21331790-0048198 & 15.87  & 14.74  & 14.17  & 13.54  & 12.00  & 11.05  & 10.90 \\ 
21331854-0051563 & 14.88  & 13.89  & 13.32  & 12.72  & 11.80  & 11.17  & 11.06 \\ 
21331948-0051034 & 15.11  & 14.12  & 13.55  & 12.93  & 12.01  & 11.38  & 11.17 \\ 
21331923-0049058 & 16.49  & 15.21  & 14.58  & 13.91  & 12.52  & 11.67  & 11.46 \\ 
21332588-0046004 & 15.42  & 14.44  & 13.89  & 13.33  & 12.40  & 11.78  & 11.59 \\ 
21333668-0051058 & 14.35  & 13.12  & 12.47  & 11.81  & 10.76  & 10.05  &  9.92 \\ 
21333520-0046089 & 14.58  & 13.36  & 12.70  & 12.04  & 11.02  & 10.34  & 10.23 \\ 
21333488-0047572 & 14.98  & 13.67  & 12.98  & 12.30  & 11.33  & 10.64  & 10.42 \\ 
21333593-0049224 & 15.31  & 13.93  & 13.18  & 12.48  & 11.49  & 10.82  & 10.60 \\ 
21333432-0051285 & 14.89  & 13.72  & 13.09  & 12.45  & 11.50  & 10.87  & 10.77 \\ 
21332531-0052511 & 15.26  & 14.11  & 13.51  & 12.90  & 11.98  & 11.38  & 11.18 \\ 
21333109-0054522 & 15.98  & 14.57  & 13.80  & 13.08  & 12.15  & 11.56  & 11.37 \\ 
21333507-0051097 & 15.94  & 14.78  & 14.14  & 13.52  & 12.63  & 12.07  & 12.02 \\
\hline
\hline
\multicolumn{8}{c}{\large{\bf M~71}}\\
\multicolumn{1}{c}{ID} & \multicolumn{1}{c}{B} & \multicolumn{1}{c}{V} & \multicolumn{1}{c}{R$_c$} & \multicolumn{1}{c}{I$_c$} &
\multicolumn{1}{c}{J} & \multicolumn{1}{c}{H} & \multicolumn{1}{c}{K} \\
\hline
19535325+1846471 & 14.02  & 12.37  & 11.42  & 10.43  &  8.92  &  8.36  &  7.93 \\ 
19534750+1846169 & 14.48  & 13.11  & 12.36  & 11.59  & 10.41  &  9.71  &  9.57 \\ 
19535150+1848059 & 13.99  & 12.32  & 11.34  & 10.38  &  9.10  &  8.30  &  8.02 \\ 
19535064+1849075 & 14.46  & 13.02  & 12.22  & 11.42  & 10.27  &  9.49  &  9.25 \\ 
19534575+1847547 & 14.21  & 12.48  & 11.45  & 10.43  &  9.12  &  8.30  &  8.02 \\ 
19534827+1848021 & 14.04  & 12.36  & 11.40  & 10.47  &  9.24  &  8.41  &  8.09 \\ 
19534656+1847441 & 14.85  & 13.63  & 12.93  & 12.21  & 11.22  & 10.56  & 10.47 \\ 
19535369+1846039 & 15.50  & 14.54  & 14.01  & 13.40  & 12.56  & 12.02  & 11.85 \\ 
19534905+1846003 & 14.73  & 13.45  & 12.75  & 12.01  & 10.91  & 10.26  & 10.15 \\ 
19534916+1846512 & 14.87  & 13.65  & 13.01  & 12.30  & 11.19  & 10.46  & 10.30 \\ 
19534178+1848384 & 15.77  & 14.55  & 13.87  & 13.17  & 12.24  & 11.58  & 11.40 \\ 
19535676+1845399 & 15.66  & 14.50  & 13.86  & 13.18  & 12.22  & 11.58  & 11.43 \\ 
19533962+1848569 & 15.74  & 14.70  & 14.13  & 13.50  & 12.50  & 11.87  & 11.74 \\ 
19533864+1847554 & 15.39  & 14.10  & 13.42  & 12.72  & 11.72  & 11.19  & 11.09 \\ 
19534615+1847261 & 14.68  & 13.21  & 12.38  & 11.55  & 10.35  &  9.60  &  9.44 \\ 
19535610+1847167$^{(c)}$ & 14.96  & 13.27  & 11.21  &  9.26  &  7.89  &  7.08  &  6.83 \\ 
19534941+1844269$^{(d)}$ & 13.88  & 12.02  & 10.71  &  9.42  &  7.96  &  7.20  &  6.96 \\ 
\hline
\end{tabular}

\smallskip
\small{
(a) After correction for the systematic offsets, according to Table~\ref{t7}\\
(b) Dropped: I outlier\\
(c) Dropped: SR variable Z Sge; B, V outlier;\\
(d) Var AN 48.1928 \citep{baade28} \hfill}\\
}
\label{t8}
\end{table}

\begin{table}
\caption{Standard synthetic photometry from SED of target stars in open clusters NGC~188 and NGC~6791$^{(a)}$}
\scriptsize{
\begin{tabular}{lrrrrrrr}
\hline
\multicolumn{8}{c}{\large{\bf NGC~188}}\\
\multicolumn{1}{c}{ID} & \multicolumn{1}{c}{B} & \multicolumn{1}{c}{V} & \multicolumn{1}{c}{R$_c$} & \multicolumn{1}{c}{I$_c$} &
\multicolumn{1}{c}{J} & \multicolumn{1}{c}{H} & \multicolumn{1}{c}{K} \\
\hline
00445253+8514055 & 13.65  & 12.34  & 11.66  & 11.01  & 10.04  &  9.41  &  9.45 \\ 
00475922+8511322 & 13.68  & 12.13  & 11.38  & 10.73  &  9.78  &  9.20  &  9.25 \\ 
00465966+8513157 & 13.87  & 12.40  & 11.71  & 11.11  & 10.25  &  9.71  &  9.67 \\ 
00453697+8515084$^{(b)}$ & 15.49  & 13.90  & 13.10  & 12.40  & 11.44  & 10.87  & 10.77 \\ 
00442946+8515093$^{(c)}$ & 15.84  & 14.28  & 13.44  & 12.73  & 11.81  & 11.27  & 11.32 \\ 
00473222+8511024$^{(d)}$ & 15.79  & 14.24  & 13.40  & 12.73  & 11.87  & 11.40  & 11.49 \\ 
00554526+8512209 & 12.25  & 10.82  & 10.06  &  9.34  &  8.31  &  7.59  &  7.54 \\ 
00463920+8523336 & 12.89  & 11.56  & 10.87  & 10.21  &  9.19  &  8.53  &  8.50 \\ 
00472975+8524140 & 14.05  & 12.91  & 12.33  & 11.74  & 10.70  & 10.11  &  9.98 \\ 
00441241+8509312 & 12.81  & 11.28  & 10.45  &  9.68  &  8.60  &  7.85  &  7.78 \\ 
00432696+8509175 & 14.17  & 13.14  & 12.61  & 12.08  & 11.25  & 10.76  & 10.59 \\ 
00471847+8519456$^{(e)}$ & 14.89  & 13.03  & 12.19  & 11.47  & 10.54  &  9.99  & 10.07 \\ 
00461981+8520086$^{(f)}$ & 14.58  & 12.50  & 11.60  & 10.88  &  9.98  &  9.43  &  9.40 \\ 
00463004+8511518$^{(g)}$ & 15.67  & 14.20  & 13.45  & 12.78  & 11.91  & 11.42  & 11.38 \\ 
00490560+8526077 & 13.89  & 12.65  & 12.02  & 11.41  & 10.48  &  9.90  &  9.97 \\ 
00420323+8520492 & 11.29  &  9.75  &  8.91  &  8.14  &  7.05  &  6.43  &  6.20 \\ 
\hline
\hline
\multicolumn{8}{c}{\large{\bf NGC~6791}}\\
\multicolumn{1}{c}{ID} & \multicolumn{1}{c}{B} & \multicolumn{1}{c}{V} & \multicolumn{1}{c}{R$_c$} & \multicolumn{1}{c}{I$_c$} &
\multicolumn{1}{c}{J} & \multicolumn{1}{c}{H} & \multicolumn{1}{c}{K} \\
\hline
19210807+3747494 & 15.29  & 13.98  & 13.33  & 12.70  & 11.66  & 11.06  & 10.87 \\ 
19204971+3743426$^{(h)}$ & 15.52  & 13.97  & 12.31  & 10.61  &  9.04  &  8.20  &  7.97 \\ 
19205259+3744281 & 15.69  & 14.10  & 13.19  & 12.32  & 11.17  & 10.45  & 10.20 \\ 
19205580+3742307 & 16.23  & 14.90  & 14.23  & 13.59  & 12.61  & 12.03  & 11.96 \\ 
19205671+3743074 & 15.97  & 14.64  & 13.96  & 13.31  & 12.32  & 11.74  & 11.66 \\ 
19210112+3742134 & 15.94  & 14.48  & 13.68  & 12.92  & 11.82  & 11.06  & 10.96 \\ 
19211606+3746462 & 15.29  & 13.74  & 12.11  & 10.45  &  8.91  &  8.10  &  7.84 \\ 
19213656+3740376 & 15.58  & 14.07  & 13.24  & 12.46  & 11.40  & 10.72  & 10.50 \\ 
19210326+3741190 & 15.72  & 14.39  & 13.69  & 13.01  & 12.03  & 11.44  & 11.36 \\ 
19213635+3739445 & 16.10  & 14.76  & 14.11  & 13.49  & 12.47  & 11.78  & 11.58 \\ 
19212437+3735402 & 15.83  & 14.46  & 13.79  & 13.15  & 12.14  & 11.45  & 11.30 \\ 
19212674+3735186 & 15.33  & 13.98  & 13.25  & 12.56  & 11.59  & 10.99  & 10.79 \\ 
19211632+3752154 & 15.22  & 13.98  & 13.32  & 12.67  & 11.71  & 11.12  & 10.93 \\ 
19211176+3752459 & 15.65  & 14.38  & 13.73  & 13.11  & 12.13  & 11.47  & 11.41 \\ 
19202345+3754578 & 14.57  & 12.80  & 11.43  &  9.63  &  7.96  &  7.08  &  6.84 \\ 
19205149+3739334 & 13.26  & 11.66  & 10.23  &  8.75  &  7.35  &  6.58  &  6.23 \\ 
19203285+3753488 & 14.97  & 13.39  & 11.67  & 10.00  &  8.46  &  7.55  &  7.19 \\ 
19200641+3744452 & 14.67  & 13.34  & 12.61  & 11.89  & 10.79  & 10.06  &  9.87 \\ 
19200882+3744317 & 15.17  & 13.62  & 11.61  &  9.67  &  7.92  &  7.07  &  6.72 \\ 
19203219+3744208$^{(i)}$ & 16.22  & 14.77  & 12.46  & 10.23  &  8.18  &  7.29  &  6.93 \\ 
\hline
\end{tabular}

\smallskip
\small{
(a) After correction for the systematic offsets, according to Table~\ref{t7}\\
(b) Dropped: B, J outlier; \\
(c) Dropped: B, J, H outlier; \\
(d) Dropped: B, J outlier; \\
(e) Dropped: B, J outlier; \\
(f) Dropped: B, J, H outlier; \\
(g) Dropped: B, J outlier;\\
(h) Dropped: I outlier; V13 - Var? \citep{demarchi07} \\
(i) V70 $\equiv$ SBG~2240: Irr var \citep{moc03}\hfill}
}
\label{t9}
\end{table}

\subsection{Stellar outliers}

It could be interesting to analyze in some detail the deviant stars in 
our $\Delta m$ clipping procedure in order to collect further clues 
about their nature. Apart from the obvious impact of
photometric errors, $3\sigma$ outliers may in fact more likely be displaying signs
of an intrinsic physical variability in their luminosity.

\begin{figure}
\centerline{
\psfig{file=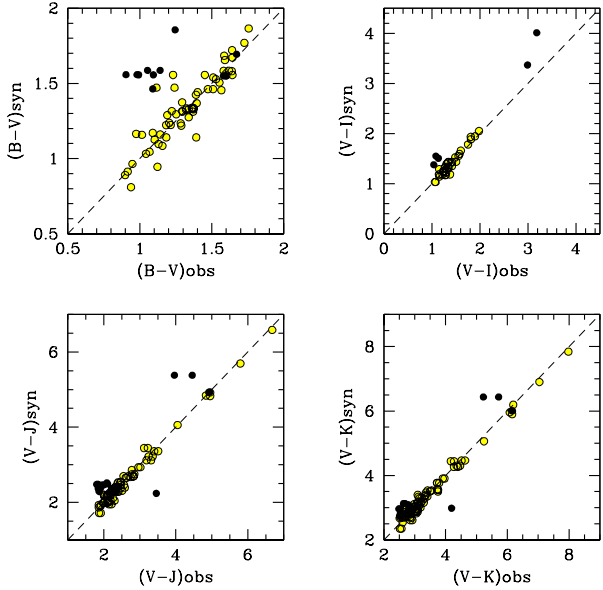,width=\hsize}
}
\caption{Color distribution of photometric outliers, according to our 3$\sigma$ clipping
procedure (see Fig.~\ref{f2}). Target location for the whole star sample in the synthetic 
vs.\ observed color planes are displayed, with dark solid dots marking the ``dropped'' 
objects (see Table~\ref{t10}).
}
\label{f3}
\end{figure}

As summarized in Table~\ref{t10}, in total 10 stars have been found to significantly 
($>3\sigma$) deviate from the literature compilations. A careful check of their identifications
on the {\sc Simbad} database indicates that at least 3 of them are known
variable (typically semir-regulars or irregulars, as expected for their
nature of late-type red giants).\footnote{Note, on the other hand, the counter-example 
of star \#2156 in NGC~6791, known as Irr variable V70 $\equiv$ SBG~2240
\citep[][]{moc03} and not a deviant in our spectroscopic observations.}
No firm conclusions can be drawn, on the contrary, for the other seven cases, although
it is evident even from a color check (see Fig.~\ref{f3}) that they have been picked up 
in an intrinsically different status with respect to previous 
data in the literature.\footnote{Curiously enough, however, one may note that 6 out of 
the 7 remaining objects are all located in NGC~188, and are both $B$ and $J$ outliers.
Both photometric bands actually cover the ``bluer'' wavelength regions of both the LRS and NICS
spectra, respectively. This coincidence might perhaps indicate some hidden problem
with the flux calibration procedure during the observation of this cluster.}
We therefore commend the stars in the list of Table~\ref{t10} as special 
candidates for further in-depth investigations for variability.

\section{Spectral energy distribution and bolometric luminosity}

The synthetic photometric catalogs obtained from the observed spectral database
had a twofold aim: firstly, this procedure allowed us to 
self-consistently match broad-band magnitudes with the inferred measure
of $m_{\rm bol}$, in order to obtain the corresponding value of the bolometric
correction; secondly, the study of the magnitude residuals with
respect to the literature data provided us with the appropriate offsets
in flux rescaling such as to ``smoothly'' connect our optical and infrared 
spectra and lead therefore to a more accurate estimate of $m_{\rm bol}$.

\begin{table}
\caption{Stellar outliers of our sample in the different photometric bands}
\scriptsize{
\begin{tabular}{lrcccccccl}
\hline 
Cluster  & ID   & \multicolumn{7}{c}{Outlier in} & Notes \\
         &      &  B & V & R$_c$ & I$_c$ & J & H & K &       \\
\hline
M15      & 180  &... &...&...  &  x  &...&...&...&...    \\ 
M71      & 4212 &  x & x &...  &...  &...&...&...& Z Sge - SRa P$\sim 175^d$\\ 
         & 5755 &... &...&...  &  x  &...&...&...& V2 $\equiv$ AN 48.1928 - Ir \\   
N188     & 567  &  x &...&...  &...  & x &...&...&...     \\
         & 652  &  x &...&...  &...  & x & x &...&...     \\
         & 1109 &  x &...&...  &...  & x &...&...&...     \\
         & 630  &  x &...&...  &...  & x &...&...&...     \\
         & 174  &  x &...&...  &...  & x & x &...&...     \\
         & 1352 &  x &...&...  &...  & x &...&...&...     \\
N6791    & 3502 &... &...&...  &  x  &...&...&...& V13 - Var? \\
\hline
   Tot   &      &  7 & 1 &  0  &  3  & 6 & 2 & 0 &        \\
\hline
\end{tabular}
}
\label{t10}
\end{table}

\begin{figure}
\centerline{
\psfig{file=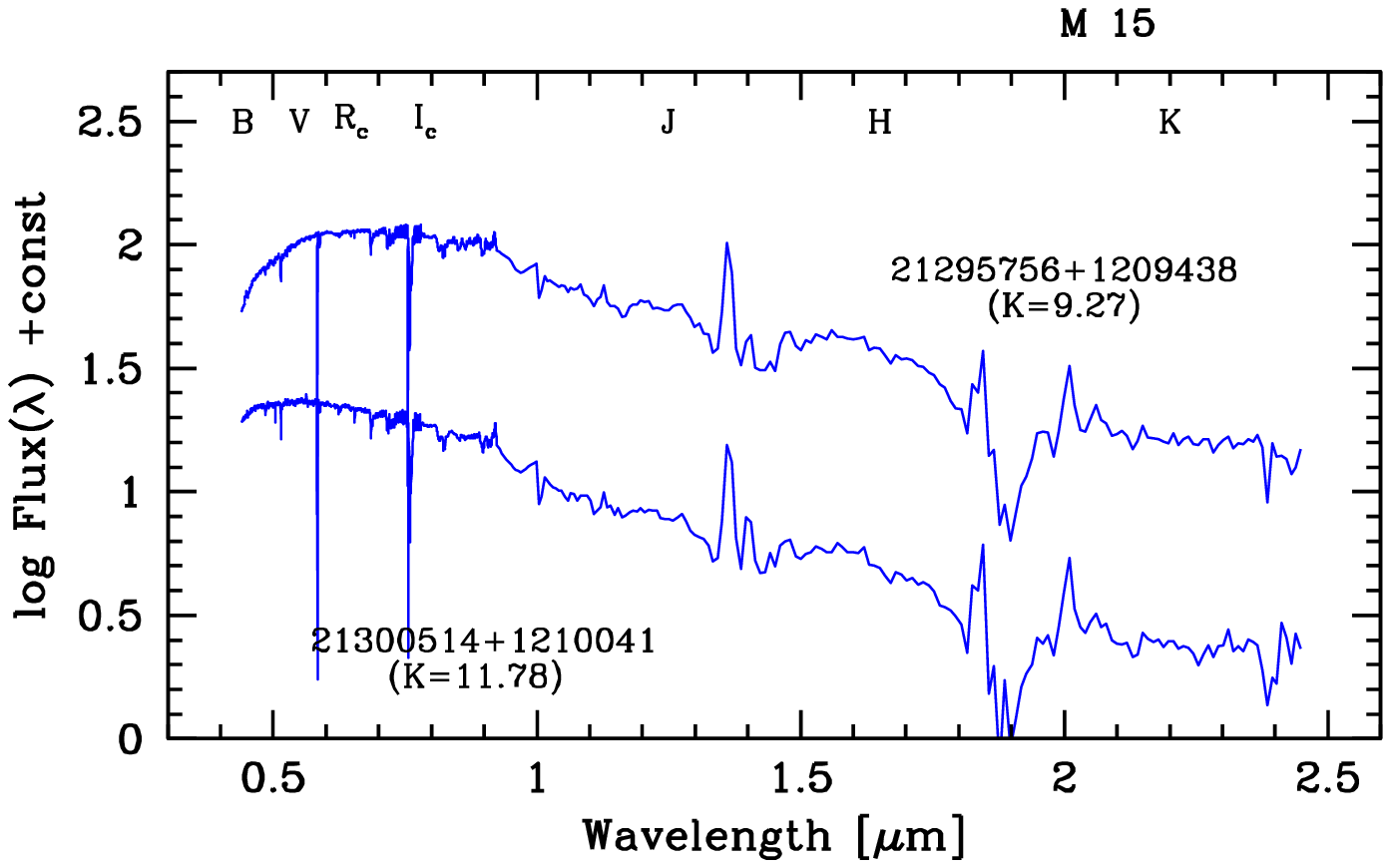,width=0.82\hsize}
} \centerline{
\psfig{file=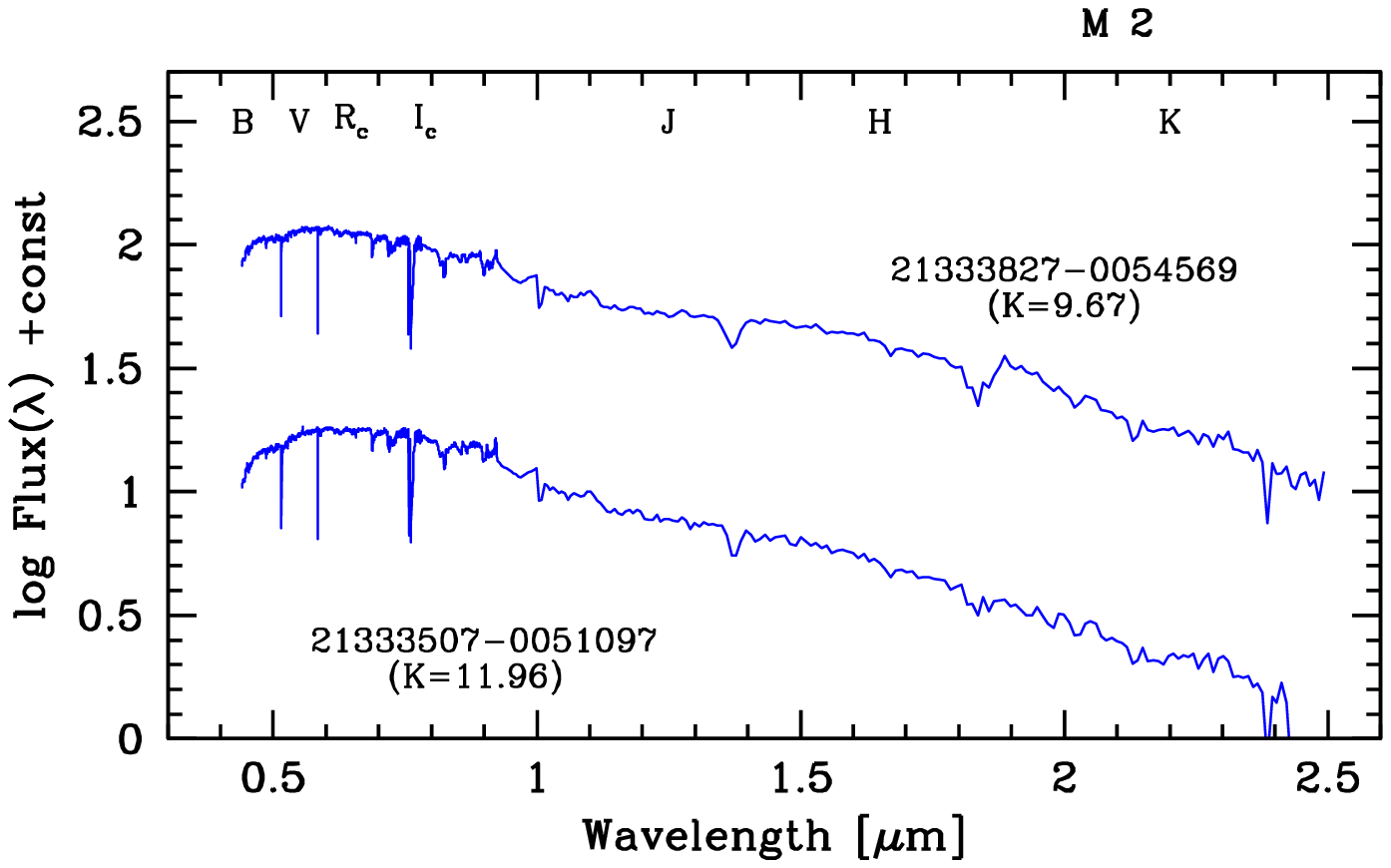,width=0.82\hsize}
} \centerline{
\psfig{file=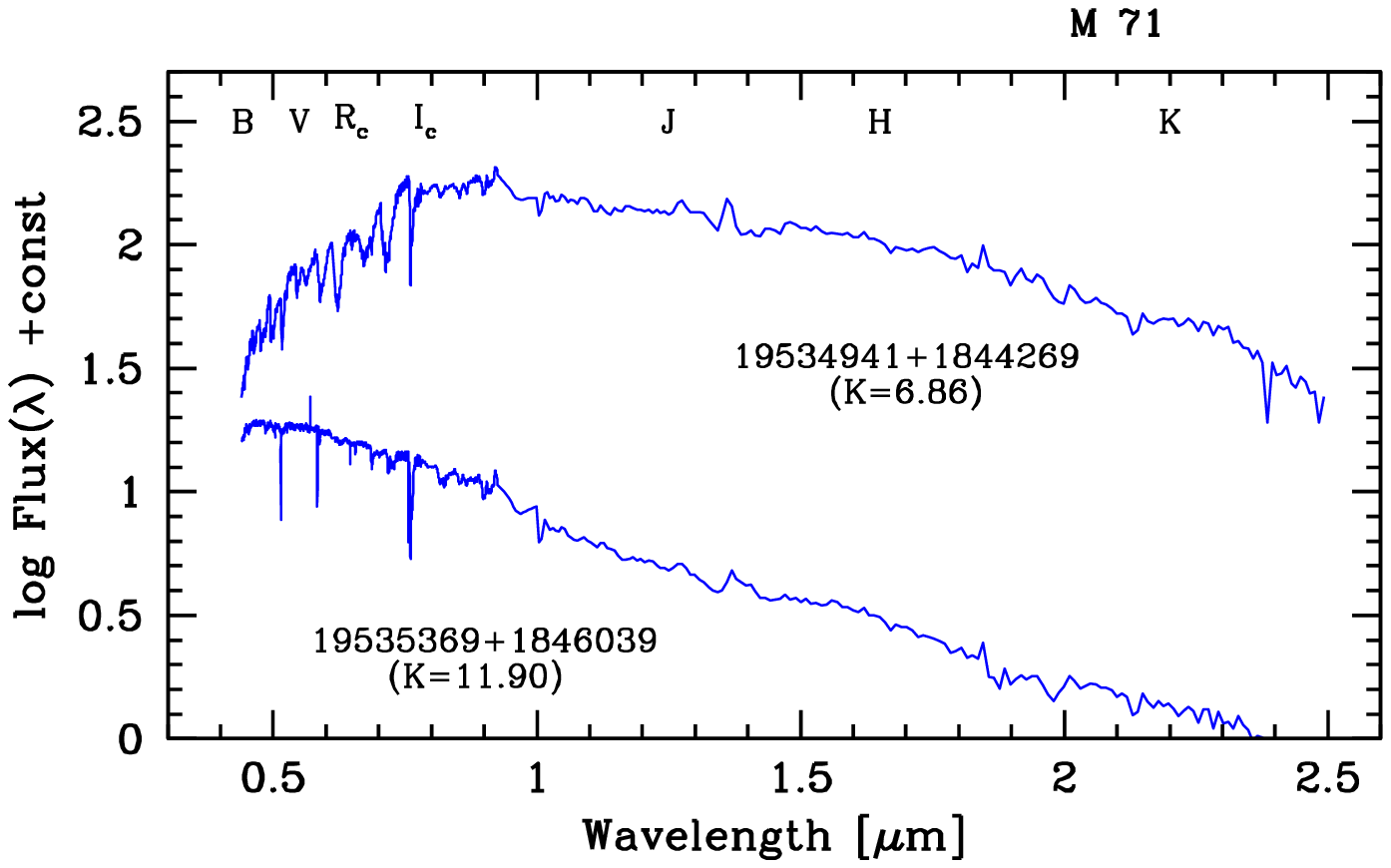,width=0.82\hsize}
} \centerline{
\psfig{file=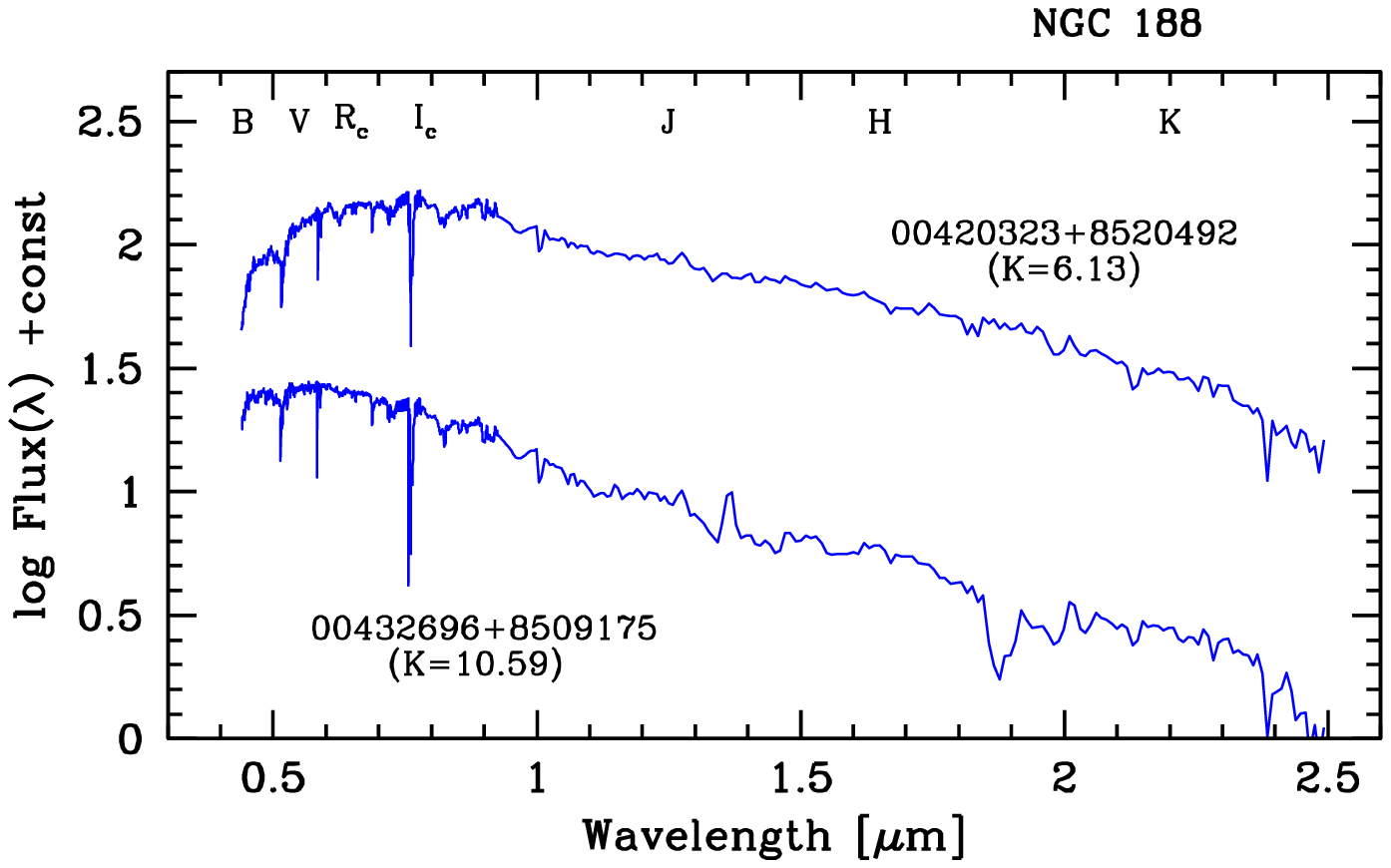,width=0.82\hsize}
} \centerline{
\psfig{file=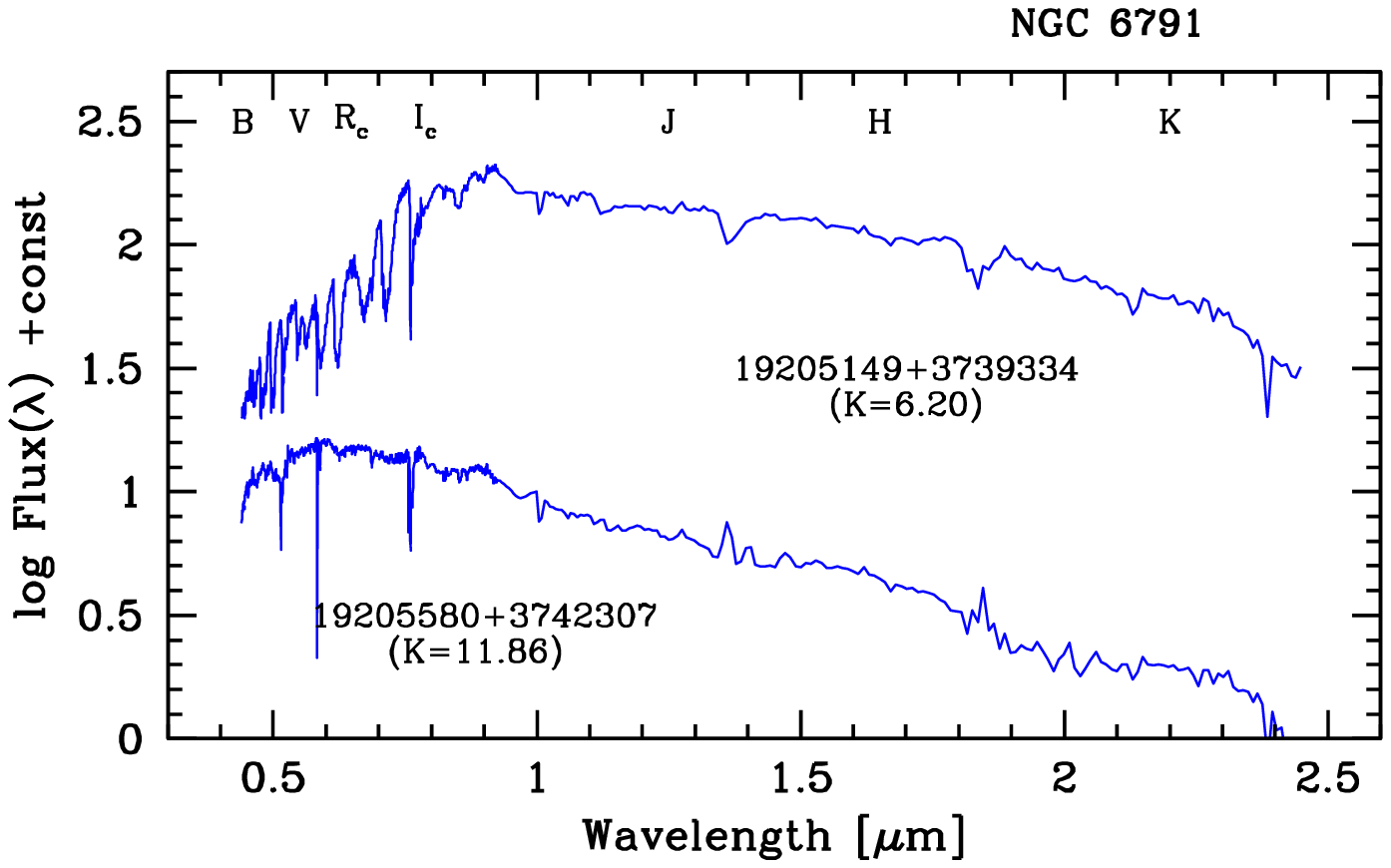,width=0.82\hsize}
}
\caption{The resulting (dereddened) SED according to optical and infrared observations
for an illustrative stellar subset of each cluster, including the brightest (and 
roughly coolest) and faintest (i.e.\ warmest) stars. Note, especially for the
M~15 stars, the strong impact of telluric water vapor bands at 1.38 and 1.88~$\mu$m.
Their variability along the observing nights prevented, in some cases, any accurate
cleaning procedure. See discussion in Sec.~3.1.
}
\label{f35}
\end{figure}

Operationally, for the latter task, we proceeded as follows.
Taking into account the individual set of $\langle Obs - Syn \rangle$
magnitude residuals, for each star in our sample we computed 
a mean optical and infrared offset ($\Delta m_{\rm LRS}$ and 
$\Delta m_{\rm NICS}$, respectively) by separately averaging the $B,V,R_c,I_c$ 
and $J,H,K$ mag residuals.
The LRS spectra and the NICS observations have then been matched  
by multiplying visual and IR fluxes by a factor $10^{-0.4\,(\Delta m_{\rm LRS})}$ 
and $10^{-0.4\,(\Delta m_{\rm NICS})}$, respectively .
Foreground reddening has been corrected relying on the standard relation  
$k(\lambda) = A(\lambda)/E(B-V)$ \citep{landolt06}, where the appropriate 
value of the color excess $E(B-V)$ is from the headers of Tables~\ref{t2} 
to \ref{t6}. In its final form, the SED is reshaped 
such as $f_o(\lambda) = f(\lambda)\, 10^{0.4\,k(\lambda)E(B-V)}$.

The LRS and NICS spectra have been connected at 8800~\AA, by smoothing the wavelength 
region between 7800 and 10000 \AA\ (in order to gain S/N, especially for LRS poor signal
at the long wavelenegth edge).  
In Fig.~\ref{f35}, we summarize our results for an illustrative set of SEDs by picking
up for each cluster the brightest (i.e.\ roughly the coolest) and faintest 
(i.e.\ warmest) stars in our sample.\footnote{For the interested reader, the entire 
spectral database is available in electronic form upon request, or directly on 
the web at the authors' web site {\tt http://www.bo.astro.it/$\sim$eps/home.html}.}
Note, from the figure, the stricking presence of the CO bump about 1.6~$\mu$m
\citep{frogel78,lancon02}, as well as the broad H$_2$O absorption bands
to which the sharper (and variable) emission of telluric water vapor superposes (see,
in particular, the case of M15 stars in the figure). This made far more difficult any accurate
cleaning procedure, as we discussed in Sec.~3.1.

\subsection{Temperature scale}

Although sampled over a wide wavelength range, SED of our stars still lacks 
the contribution of ultraviolet and far-infrared luminosity. Clearly, a safe assessment 
of this contribution is mandatory to lead to a confident measure of the bolometric magnitude.
As the amount of energy released outside the spectral window of our observations
critically depends on stellar temperature, our task to compute $BC$ 
requires in fact a parallel calibration of $T_{\rm eff}$ in the range 
of our red giant stars.

Among the many outstanding efforts in this direction, we have to recall the works
of \citet{flower75}, \citet{bessell79}, \citet{black80}, \citet{ridgway80}, 
\citet{bessell98}, \citet{houd00}, \citet{vdb03}, 
\citet{bertone04} and \citet{worthey06}. 
In their exhaustive analysis, \citet{alonso99} provided
an accurate analytical set of fitting functions, that calibrate stellar effective 
temperature vs.\ Johnson/Cousins broad-band colors. The \citet{alonso99}
calibration relies on the IRFM estimate of stellar surface brightness, 
and considers stars of spectral type K5 or earlier,
spanning a wide metallicity range ($-3.0 \la [Fe/H] \la +0.2$).
Within this range, the Authors claim an internal accuracy in the definition of
$T_{\rm eff}$ better than 5\%. As a further important result of their work,
some colors, like $(V-I)$, $(V-L')$, $(J-K)$ and $(I-K)$ are found to be
fair tracers of temperature, almost independently from stellar metallicity.

The \citet{alonso99} calibration, however, strictly applies only to stars warmer 
than $\sim 4000$~K, while our stellar sample definitely spans a wider color range.
This is certainly the case, for instance, of the brightest giant
stars in NGC~6791, too (infra)red to match the Alonso et al.\ fitting
functions. For these cases one could rely on the wider validity range of 
the $(B-V)$ calibration, although the advantage may only be a nominal one
as any optical color, like $(B-V)$ tends naturally to saturate when moving to  
$T_{\rm eff} \la 4000$~K \citep[][see also Fig.~2 in \citealp{alonso99}]{johnson66}.

Considering the whole set of the Alonso et al.\ fitting functions, we eventually
chose four reference colors to assess the value of effective temperature for our stars.
Two colors, namely $(B-V)$ and $(J-K)$ are entirely comprised within the LRS
and NICS spectral branches, respectively, and they can therefore ostensibly probe
the shape of SED in a more self-consistent way. To these two colors we also added
$(V-I_c)$ and $(V-K)$, as they provided a check of our flux calibration bridging 
the optical and infrared regions of the spectra.

\begin{table*}
\caption{Inferred temperatures, bolometric magnitude and bolometric corrections for target stars in globular clusters M~71, M~15, and M~2}
\scriptsize{
\begin{tabular}{rccccllllllrrr}
\hline
\multicolumn{13}{c}{\large{\bf M~15}}\\
\multicolumn{1}{c}{ID} & \multicolumn{1}{c}{(B-V)$_o$} & \multicolumn{1}{c}{(V-I$_c$)$_o$} & \multicolumn{1}{c}{(V-K)$_o$} & \multicolumn{1}{c}{(J-K)$_o$} &
\multicolumn{1}{c}{T$_{\rm BV}$} & \multicolumn{1}{c}{T$_{\rm VI}$} & \multicolumn{1}{c}{T$_{\rm VK}$} & \multicolumn{1}{c}{T$_{\rm JK}$} &  & $\langle$T$\rangle$ &
\multicolumn{1}{c}{Bol$_o$} & \multicolumn{1}{c}{BC$_V$} & \multicolumn{1}{c}{BC$_K$}\\
 &  &  & &  & \multicolumn{1}{c}{$^o$K} & \multicolumn{1}{c}{$^o$K} & \multicolumn{1}{c}{$^o$K} & \multicolumn{1}{c}{$^o$K} &  & $^o$K &  & &  \\
\hline
21300002+1209182 & 0.71 & 0.87 & 2.08 & 0.58 & 5043 & 5039 & 5019 & 4748 &  & 4962 & 13.742 & --0.25 & 1.83 \\
21295705+1208531 & 0.85 & 1.01 & 2.34 & 0.62 & 4721 & 4711 & 4723 & 4619 &  & 4694 & 12.697 & --0.34 & 1.99 \\
21295532+1210327 & 0.81 & 1.03 & 2.44 & 0.73 & 4777 & 4669 & 4618 & 4307 &  & 4593 & 13.754 & --0.37 & 2.07 \\
21300090+1208571 & 1.04 & 1.27 & 3.13 & 0.96 & 4474 & 4242 & 4117 & 3806 &  & 4160 & 11.918 & --0.67 & 2.45 \\
21295473+1208592 & 0.99 & 1.09 & 2.58 & 0.62 & 4536 & 4549 & 4506 & 4619 &  & 4552 & 13.032 & --0.46 & 2.12 \\
21300461+1210327 & 1.06 & 1.07 & 2.60 & 0.83 & 4449 & 4588 & 4490 & 4068 &  & 4399 & 13.148 & --0.39 & 2.20 \\
21295560+1212422 & 1.03 & 1.07 & 2.64 & 0.81 & 4486 & 4588 & 4457 & 4113 &  & 4411 & 12.733 & --0.42 & 2.22 \\
21300514+1210041 & 0.79 & 0.92 & 2.40 & 0.76 & 4827 & 4915 & 4660 & 4231 &  & 4658 & 13.674 & --0.33 & 2.07 \\
21295836+1209020 & 0.87 & 1.02 & 2.55 & 0.75 & 4694 & 4690 & 4532 & 4256 &  & 4543 & 13.102 & --0.42 & 2.13 \\
21295618+1210179 & 1.03 & 1.25 & 3.07 & 0.89 & 4486 & 4272 & 4153 & 3941 &  & 4213 & 11.934 & --0.65 & 2.42 \\
21295739+1209056 & 0.96 & 1.09 & 2.53 & 0.57 & 4574 & 4549 & 4549 & 4782 &  & 4614 & 13.068 & --0.45 & 2.07 \\
21300097+1210375 & 0.96 & 1.06 & 2.44 & 0.66 & 4574 & 4608 & 4618 & 4499 &  & 4575 & 13.135 & --0.41 & 2.03 \\
21300431+1210561 & 1.07 & 1.18 & 2.67 & 0.70 & 4437 & 4386 & 4433 & 4387 &  & 4411 & 12.798 & --0.48 & 2.18 \\
21301049+1210061 & 0.99 & 1.11 & 2.74 & 0.85 & 4536 & 4511 & 4378 & 4024 &  & 4362 & 12.586 & --0.46 & 2.27 \\
21300739+1210330 & 1.07 & 1.22 & 2.68 & 0.71 & 4437 & 4320 & 4425 & 4360 &  & 4386 & 13.252 & --0.49 & 2.19 \\
21300569+1210156 & 0.81 & 0.98 & 2.34 & 0.66 & 4777 & 4776 & 4723 & 4499 &  & 4694 & 13.472 & --0.33 & 2.01 \\
21300553+1208553 & 0.88 & 1.07 & 2.92 & 0.80 & 4680 & 4588 & 4249 & 4136 &  & 4413 & 14.012 & --0.52 & 2.40 \\
21295756+1209438 & 1.25 & 1.28 & 2.86 & 0.73 & 4230 & 4227 & 4291 & 4307 &  & 4264 & 11.556 & --0.58 & 2.27 \\
21295082+1211301 & 0.99 & 1.09 & 2.59 & 0.69 & 4536 & 4549 & 4498 & 4414 &  & 4499 & 12.778 & --0.43 & 2.15 \\
21295716+1209175 & 0.83 & 1.03 & 2.56 & 0.73 & 4749 & 4669 & 4523 & 4307 &  & 4562 & 12.303 & --0.42 & 2.14 \\
\hline
\hline								  
\multicolumn{13}{c}{\large{\bf M~2}}\\
\multicolumn{1}{c}{ID} & \multicolumn{1}{c}{(B-V)$_o$} & \multicolumn{1}{c}{(V-I$_c$)$_o$} & \multicolumn{1}{c}{(V-K)$_o$} & \multicolumn{1}{c}{(J-K)$_o$} &
\multicolumn{1}{c}{T$_{\rm BV}$} & \multicolumn{1}{c}{T$_{\rm VI}$} & \multicolumn{1}{c}{T$_{\rm VK}$} & \multicolumn{1}{c}{T$_{\rm JK}$} &  & $\langle$T$\rangle$ &
\multicolumn{1}{c}{Bol$_o$} & \multicolumn{1}{c}{BC$_V$} & \multicolumn{1}{c}{BC$_K$}\\
 &  &  & &  & \multicolumn{1}{c}{$^o$K} & \multicolumn{1}{c}{$^o$K} & \multicolumn{1}{c}{$^o$K} & \multicolumn{1}{c}{$^o$K} &  & $^o$K &  & &  \\
\hline
21333827-0054569 & 0.93 & 1.04 & 2.83 & 0.79 & 4640 & 4639 & 4312 & 4157 &  & 4437 & 11.990 & --0.54 & 2.28 \\
21333095-0052154 & 1.00 & 1.05 & 2.65 & 0.69 & 4540 & 4618 & 4447 & 4412 &  & 4504 & 13.025 & --0.47 & 2.18 \\
21332468-0044252 & 1.00 & 1.07 & 2.55 & 0.62 & 4540 & 4579 & 4530 & 4617 &  & 4566 & 14.030 & --0.43 & 2.11 \\
21331771-0047273 & 1.23 & 1.26 & 2.95 & 0.75 & 4251 & 4250 & 4230 & 4254 &  & 4246 & 12.190 & --0.64 & 2.30 \\
21331723-0048171 & 1.34 & 1.35 & 3.18 & 0.83 & 4110 & 4122 & 4090 & 4066 &  & 4097 & 12.721 & --0.75 & 2.42 \\
21331790-0048198 & 1.07 & 1.10 & 3.68 & 1.07 &      & 4521 & {\it 3912}& 3618 &  & 4017 & 13.584 & --0.97 & 2.71 \\
21331854-0051563 & 0.93 & 1.07 & 2.67 & 0.71 & 4640 & 4579 & 4431 & 4358 &  & 4502 & 13.231 & --0.47 & 2.19 \\
21331948-0051034 & 0.93 & 1.09 & 2.79 & 0.81 & 4640 & 4540 & 4340 & 4111 &  & 4408 & 13.428 & --0.51 & 2.28 \\
21331923-0049058 & 1.22 & 1.20 & 3.59 & 1.03 &      & 4345 & {\it 3950}& 3682 &  & 3992 & 14.083 & --0.94 & 2.64 \\
21332588-0046004 & 0.92 & 1.01 & 2.69 & 0.78 & 4655 & 4701 & 4416 & 4181 &  & 4488 & 13.814 & --0.44 & 2.25 \\
21333668-0051058 & 1.17 & 1.21 & 3.04 & 0.81 & 4314 & 4328 & 4173 & 4111 &  & 4232 & 12.271 & --0.66 & 2.37 \\
21333520-0046089 & 1.16 & 1.22 & 2.97 & 0.76 & 4326 & 4312 & 4217 & 4229 &  & 4271 & 12.528 & --0.65 & 2.32 \\
21333488-0047572 & 1.25 & 1.27 & 3.09 & 0.88 & 4215 & 4235 & 4143 & 3959 &  & 4138 & 12.813 & --0.67 & 2.41 \\
21333593-0049224 & 1.32 & 1.35 & 3.17 & 0.86 & 4133 & 4122 & 4096 & 4001 &  & 4088 & 13.002 & --0.74 & 2.42 \\
21333432-0051285 & 1.11 & 1.17 & 2.79 & 0.70 & 4391 & 4395 & 4340 & 4384 &  & 4378 & 12.980 & --0.55 & 2.23 \\
21332531-0052511 & 1.09 & 1.11 & 2.77 & 0.77 & 4417 & 4502 & 4355 & 4205 &  & 4370 & 13.414 & --0.51 & 2.26 \\
21333109-0054522 & 1.35 & 1.39 & 3.04 & 0.75 & 4098 & 4070 & 4173 & 4254 &  & 4149 & 13.675 & --0.71 & 2.33 \\
21333507-0051097 & 1.10 & 1.16 & 2.60 & 0.58 & 4404 & 4412 & 4488 & 4746 &  & 4512 & 14.102 & --0.49 & 2.10 \\
\hline
\hline
\multicolumn{13}{c}{\large{\bf M~71}}\\
\multicolumn{1}{c}{ID} & \multicolumn{1}{c}{(B-V)$_o$} & \multicolumn{1}{c}{(V-I$_c$)$_o$} & \multicolumn{1}{c}{(V-K)$_o$} & \multicolumn{1}{c}{(J-K)$_o$} &
\multicolumn{1}{c}{T$_{\rm BV}$} & \multicolumn{1}{c}{T$_{\rm VI}$} & \multicolumn{1}{c}{T$_{\rm VK}$} & \multicolumn{1}{c}{T$_{\rm JK}$} &  & $\langle$T$\rangle$ &
\multicolumn{1}{c}{Bol$_o$} & \multicolumn{1}{c}{BC$_V$} & \multicolumn{1}{c}{BC$_K$}\\
 &  &  & &  & \multicolumn{1}{c}{$^o$K} & \multicolumn{1}{c}{$^o$K} & \multicolumn{1}{c}{$^o$K} & \multicolumn{1}{c}{$^o$K} &  & $^o$K &  & &  \\
\hline
19535325+1846471 & 1.39 & 1.54 & 3.75 & 0.86 &      & 3908 & {\it 3881}& 3999 &  & 3929 & 10.455 & --1.14 & 2.61 \\ 
19534750+1846169 & 1.11 & 1.12 & 2.85 & 0.71 & 4466 & 4496 & 4307 & 4355 &  & 4406 & 11.760 & --0.57 & 2.28 \\ 
19535150+1848059 & 1.41 & 1.54 & 3.61 & 0.95 &      & 3908 & {\it 3938}& 3821 &  & 3889 & 10.522 & --1.02 & 2.59 \\ 
19535064+1849075 & 1.18 & 1.20 & 3.08 & 0.89 & 4361 & 4356 & 4160 & 3937 &  & 4204 & 11.577 & --0.67 & 2.42 \\ 
19534575+1847547 & 1.47 & 1.65 & 3.77 & 0.97 &      &	   & {\it 3873}& 3784 &  & 3828 & 10.554 & --1.15 & 2.62 \\ 
19534827+1848021 & 1.42 & 1.49 & 3.58 & 1.02 &      & 3961 & {\it 3951}& 3697 &  & 3870 & 10.606 & --0.98 & 2.60 \\ 
19534656+1847441 & 0.96 & 1.02 & 2.47 & 0.62 & 4710 & 4694 & 4600 & 4614 &  & 4654 & 12.438 & --0.42 & 2.06 \\ 
19535369+1846039 & 0.70 & 0.74 & 2.00 & 0.58 & 5289 & 5410 & 5096 & 4742 &  & 5134 & 13.556 & --0.21 & 1.79 \\ 
19534905+1846003 & 1.02 & 1.04 & 2.61 & 0.63 & 4609 & 4653 & 4486 & 4583 &  & 4583 & 12.209 & --0.47 & 2.15 \\ 
19534916+1846512 & 0.96 & 0.95 & 2.66 & 0.76 & 4710 & 4849 & 4447 & 4227 &  & 4558 & 12.434 & --0.44 & 2.22 \\ 
19534178+1848384 & 0.96 & 0.98 & 2.46 & 0.71 & 4710 & 4781 & 4610 & 4355 &  & 4614 & 13.405 & --0.37 & 2.09 \\ 
19535676+1845399 & 0.90 & 0.92 & 2.38 & 0.66 & 4815 & 4920 & 4688 & 4494 &  & 4729 & 13.379 & --0.35 & 2.04 \\ 
19533962+1848569 & 0.78 & 0.80 & 2.27 & 0.63 & 5082 & 5234 & 4800 & 4583 &  & 4925 & 13.636 & --0.29 & 1.98 \\ 
19533864+1847554 & 1.03 & 0.98 & 2.32 & 0.50 & 4593 & 4781 & 4748 & 5029 &  & 4788 & 12.980 & --0.34 & 1.98 \\ 
19534615+1847261 & 1.21 & 1.26 & 3.08 & 0.78 & 4318 & 4260 & 4160 & 4178 &  & 4229 & 11.728 & --0.71 & 2.38 \\ 
19534941+1844269 & 1.60 & 2.20 & 4.37 & 0.87 &      &	   & {\it 3676}& 3978 &  & 3827 & 9.541  & --1.70 & 2.67 \\ 
\hline								       
\end{tabular}
}
\label{t11}
\end{table*}

\begin{figure}
\centerline{
\psfig{file=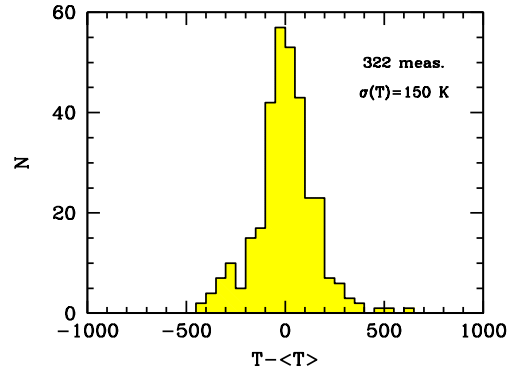,width=0.85\hsize}
}
\caption{Histogram of temperature difference for all the data reported
in Tables~\ref{t8} and \ref{t9} (cols.\ 6 to 9) with respect to the adopted
mean estimate ($\langle T \rangle$ of col.\ 10). A total of 322 entries
are available for the whole stellar sample. The resulting distribution gives
a direct measure of the internal uncertainty of our temperature scale,
amounting of $\sigma(T_{\rm eff}) = \pm 150$~K for the standard individual
estimate.
}
\label{f6}
\end{figure}

\begin{table*}
\caption{Inferred temperatures, bolometric magnitude and bolometric corrections for target stars in open clusters NGC~188 and NGC~6791}
\scriptsize{
\begin{tabular}{rccccllllllrrr}
\hline
\multicolumn{13}{c}{\large{\bf NGC~188}}\\
\multicolumn{1}{c}{ID} & \multicolumn{1}{c}{(B-V)$_o$} & \multicolumn{1}{c}{(V-I$_c$)$_o$} & \multicolumn{1}{c}{(V-K)$_o$} & \multicolumn{1}{c}{(J-K)$_o$} &
\multicolumn{1}{c}{T$_{\rm BV}$} & \multicolumn{1}{c}{T$_{\rm VI}$} & \multicolumn{1}{c}{T$_{\rm VK}$} & \multicolumn{1}{c}{T$_{\rm JK}$} &  & $\langle$T$\rangle$ &
\multicolumn{1}{c}{Bol$_o$} & \multicolumn{1}{c}{BC$_V$} & \multicolumn{1}{c}{BC$_K$}\\
 &  &  & &  & \multicolumn{1}{c}{$^o$K} & \multicolumn{1}{c}{$^o$K} & \multicolumn{1}{c}{$^o$K} & \multicolumn{1}{c}{$^o$K} &  & $^o$K &  & &  \\
\hline
00445253+851405 & 1.22 & 1.20 & 2.66 & 0.55 & 4398 & 4354 & 4470 & 4854 &  & 4519 & 11.552 & --0.53 & 2.13 \\ 
00475922+851132 & 1.46 & 1.27 & 2.65 & 0.49 & 4040 & 4243 & 4478 & 5078 &  & 4460 & 11.311 & --0.56 & 2.09 \\ 
00465966+851315 & 1.38 & 1.16 & 2.50 & 0.54 & 4153 & 4422 & 4602 & 4889 &  & 4516 & 11.701 & --0.44 & 2.06 \\ 
00554526+851220 & 1.34 & 1.35 & 3.05 & 0.73 & 4211 & 4129 & 4201 & 4309 &  & 4212 &  9.842 & --0.72 & 2.33 \\ 
00463920+852333 & 1.24 & 1.22 & 2.83 & 0.65 & 4366 & 4321 & 4344 & 4530 &  & 4390 & 10.706 & --0.60 & 2.23 \\ 
00472975+852414 & 1.05 & 1.04 & 2.70 & 0.68 & 4695 & 4650 & 4439 & 4444 &  & 4557 & 12.166 & --0.49 & 2.21 \\ 
00441241+850931 & 1.44 & 1.47 & 3.27 & 0.78 & 4068 & 3981 & 4078 & 4184 &  & 4078 & 10.165 & --0.86 & 2.41 \\ 
00432696+850917 & 0.94 & 0.93 & 2.32 & 0.62 & 4909 & 4893 & 4772 & 4621 &  & 4799 & 12.578 & --0.31 & 2.02 \\ 
00490560+852607 & 1.15 & 1.11 & 2.45 & 0.47 & 4516 & 4513 & 4646 & 5159 &  & 4708 & 11.945 & --0.45 & 2.00 \\ 
00420323+852049 & 1.45 & 1.48 & 3.32 & 0.81 & 4054 & 3970 & 4052 & 4114 &  & 4048 &  8.628 & --0.87 & 2.46 \\ 
\hline
\hline
\multicolumn{13}{c}{\large{\bf NGC~6791}}\\
\multicolumn{1}{c}{ID} & \multicolumn{1}{c}{(B-V)$_o$} & \multicolumn{1}{c}{(V-I$_c$)$_o$} & \multicolumn{1}{c}{(V-K)$_o$} & \multicolumn{1}{c}{(J-K)$_o$} &
\multicolumn{1}{c}{T$_{\rm BV}$} & \multicolumn{1}{c}{T$_{\rm VI}$} & \multicolumn{1}{c}{T$_{\rm VK}$} & \multicolumn{1}{c}{T$_{\rm JK}$} &  & $\langle$T$\rangle$ &
\multicolumn{1}{c}{Bol$_o$} & \multicolumn{1}{c}{BC$_V$} & \multicolumn{1}{c}{BC$_K$}\\
 &  &  & &  & \multicolumn{1}{c}{$^o$K} & \multicolumn{1}{c}{$^o$K} & \multicolumn{1}{c}{$^o$K} & \multicolumn{1}{c}{$^o$K} &  & $^o$K &  & &  \\
\hline
19210807+3747494 & 1.19 & 1.09 & 2.79 & 0.73 & 4554 & 4544 & 4397 & 4304 &  & 4450 & 13.100 & --0.52 & 2.27 \\
19205259+3744281 & 1.47 & 1.59 & 3.58 & 0.91 &      & 3844 & {\it 3953}& 3898 &  & 3898 & 12.720 & --1.02 & 2.56 \\
19205580+3742307 & 1.21 & 1.12 & 2.62 & 0.59 & 4518 & 4488 & 4527 & 4712 &  & 4561 & 14.053 & --0.48 & 2.13 \\
19205671+3743074 & 1.21 & 1.14 & 2.66 & 0.60 & 4518 & 4451 & 4495 & 4679 &  & 4536 & 13.773 & --0.50 & 2.15 \\
19210112+3742134 & 1.34 & 1.37 & 3.20 & 0.80 & 4294 & 4098 & 4138 & 4133 &  & 4166 & 13.322 & --0.80 & 2.40 \\
19211606+3746462 & 1.43 & 3.10 & 5.58 & 1.01 &      &	   & {\it 3418}& 3715 &  & 3566 & 10.646 & --2.73 & 2.85 \\
19213656+3740376 & 1.39 & 1.42 & 3.25 & 0.84 & 4214 & 4035 & 4111 & 4043 &  & 4101 & 12.902 & --0.81 & 2.44 \\
19210326+3741190 & 1.21 & 1.19 & 2.71 & 0.61 & 4518 & 4365 & 4456 & 4647 &  & 4496 & 13.493 & --0.53 & 2.17 \\
19213635+3739445 & 1.22 & 1.08 & 2.86 & 0.83 & 4499 & 4563 & 4348 & 4065 &  & 4369 & 13.867 & --0.53 & 2.33 \\
19212437+3735402 & 1.25 & 1.12 & 2.84 & 0.78 & 4446 & 4488 & 4361 & 4180 &  & 4369 & 13.561 & --0.54 & 2.30 \\
19212674+3735186 & 1.23 & 1.23 & 2.87 & 0.73 & 4482 & 4299 & 4341 & 4279 &  & 4350 & 13.034 & --0.58 & 2.29 \\
19211632+3752154 & 1.12 & 1.12 & 2.73 & 0.72 & 4686 & 4488 & 4441 & 4330 &  & 4486 & 13.119 & --0.50 & 2.23 \\
19211176+3752459 & 1.15 & 1.08 & 2.65 & 0.66 & 4629 & 4563 & 4503 & 4496 &  & 4548 & 13.538 & --0.48 & 2.17 \\
19202345+3754578 & 1.65 & 2.98 & 5.64 & 1.06 &      &	   & {\it 3407}& 3633 &  & 3520 & 9.718  & --2.72 & 2.92 \\
19205149+3739334 & 1.48 & 2.72 & 5.11 & 1.06 &      &	   & {\it 3504}& 3633 &  & 3568 & 9.043  & --2.25 & 2.85 \\
19203285+3753488 & 1.46 & 3.20 & 5.88 & 1.21 &      &	   & {\it 3368}& 3416 &  & 3392 & 10.107 & --2.92 & 2.96 \\
19200641+3744452 & 1.21 & 1.26 & 3.15 & 0.86 & 4518 & 4253 & 4166 & 4000 &  & 4234 & 12.259 & --0.72 & 2.43 \\
19200882+3744317 & 1.43 & 3.76 & 6.58 & 1.14 &      &	   & {\it 3260}& 3512 &  & 3386 & 9.664  & --3.59 & 2.99 \\
19203219+3744208 & 1.33 & 4.35 & 7.52 & 1.19 &      &	   & {\it 3120}& 3443 &  & 3282 & 9.934  & --4.47 & 3.05 \\
\hline
\end{tabular}
}
\label{t12}
\end{table*}

Dereddened colors for each stars in our sample provided eventually a set of nominal 
values of $T_{\rm eff}$, by entering the appropriate fitting functions.
The ``allowed'' values of $T_{\rm eff}$ (i.e.\ if comprised within the boundary 
limits of the adopted calibration functions) were then averaged, deriving the mean
fiducial value of the effective temperature,  reported in Table~\ref{t11} and \ref{t12}
(column~10).
In case of just one $T_{\rm eff}$ estimate (typically from $(J-K)$ color) we
also added the $(V-K)$ output (reported in italics in the tables) trusting on a 
fairly smooth trend of the \citet{alonso99} calibration for this color, 
when extrapolated to cooler temperatures (see Fig.~8 and Fig.~10 therein).

Once combining the different temperature estimates from the four reference colors 
in our analysis, we report in Fig.~\ref{f6} the resulting $T-\langle T\rangle$ 
distribution, considering the whole set of 322 individual residuals.
The figure confirms that an unbiased estimate of $T_{\rm eff}$ may eventually
be achieved with our procedure, within a $\pm 150$~K uncertainty on the standard 
measure. As, typically 2-4 useful 
temperature estimates are available from the colors of each star (see, again,
Table~\ref{t11} and \ref{t12}), we may expect final $T_{\rm eff}$ values 
for our sample to be assessed within a 70-100~K (i.e. 1-3\%) internal uncertainty.

\subsection{Toward m$_{\rm bol}$}

The fiducial effective temperature, as reported in col.~10
of Table~\ref{t11} and \ref{t12}, provided the reference quantity to constrain the
unsampled fraction of stellar luminosity, outside the wavelength
limits of our spectral observations. No univocal procedure can be
devised to effectively tackle this problem; from one hand, in fact,
both the ultraviolet and mid- and far-infrared stellar emission 
can in principle be modulated by a number of different mechanisms
(mass loss and stellar winds, or circumstellar gas and dust lanes
thermalizing ultraviolet and optical photons, photospheric spots,
pulsating variability etc.). On the other hand, one would better like 
to proceed with a straight heuristic approach, such as to 
self-consistently size up the amount of ``overflown'' luminosity
and decide the accuracy level in its correction procedure, according
to an {\it ``ex-post''} analysis of the results.

On this line, we therefore decided to proceed in the most straightforward way
for each star, by extrapolating its observed SED to both ultraviolet and 
infrared windows by means of two black-body branches, of appropriate
(fixed) temperature $\langle$T$\rangle$ as in Tables~\ref{t11} and \ref{t12}.
The two spectral branches have been {\it separately} rescaled to the (dereddened) 
flux values of the observed SED by setting the boundary wavelengths 
respectively at 4000 and 22500~\AA; the integrated luminosity has then
been computed within the three relevant regions of each stellar SED,
identifying the ultraviolet contribution $l_{\rm UV}$ (between $0 \le \lambda \le 4000$~\AA)
an optical/mid-infrared luminosity $l_{\rm obs}$ ($4000 \le \lambda \le 22500$~\AA)
and a far-infrared contribution $l_{\rm FIR}$ (longward of 2.25~$\mu$m). 
For comparison, the same excercise has 
been repeated for a straight black-body spectral distribution exploring the
luminosity fraction emitted shortward of $\lambda \le 4000$~\AA\ and longward
of $\lambda \ge 22500$~\AA\ along the temperature range of our sample.

\begin{figure}
\centerline{
\psfig{file=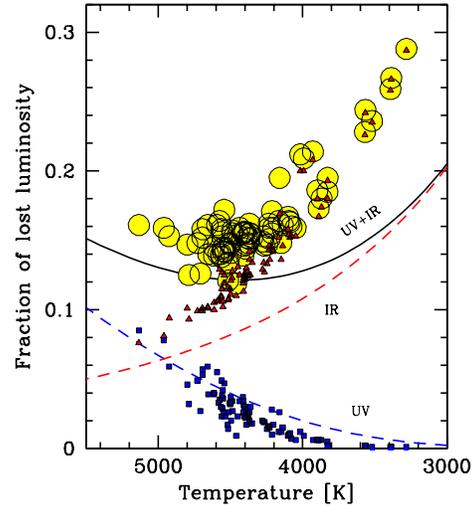,width=0.75\hsize}
}
\caption{Estimated fraction of unsampled stellar luminosity for the 
stars in our sample (big solid dots). The relative contribution to
stellar bolometric luminosity from lost emission at short (i.e.\ for 
$\lambda \le 4000$~\AA, small square markers on the plot) and long (i.e.\ for 
$\lambda \ge 2.25\mu$m, small triangles) wavelength is sized up by extrapolating
the observed SED with two black-body (BB) ``wings'' at fixed $\langle$T$\rangle$,
as from col.\ 10 of Table~\ref{t11} and \ref{t12}.
The same excercise is carried out for a full BB spectrum along the
5500-3000~K temperature range (dashed lines labelled ``UV'' and ``IR'' for
the short and long wavelength contribution, respectively, together with their
summed contribution, as in the solid line). Compared to a plain BB case, note
that real stars at cool temperatures display a brighter IR luminosity.
}
\label{f7}
\end{figure}

Our results are summarized in Fig.~\ref{f7}. Compared to the black-body 
aproximation, real stars are brighter at longer wavelength and slightly
fainter, on the contrary, at UV wavelength. 
In total, one sees from Fig.~\ref{f7} that the fraction of
``lost'' luminosity, namely $F_l = (l_{\rm bol} - l_{\rm obs})/l_{\rm bol}$, turns to be 
about 15\% for the bulk of red giants in our sample; this figure can however 
quickly raise with decreasing temperature, and about 1/3 of bolometric 
luminosity might in fact be ``stored'' at FIR wavelengths.
Within these limits, and accounting for the 70-100~K internal uncertainty of 
our temperature scale, one sees from the Fig.~\ref{f7} that $m_{\rm bol}$ 
can be secured for our sample stars within a few 0.01~mag 
uncertainty.\footnote{The claimed $m_{\rm bol}$  uncertainty simply derives 
as $\sigma \sim \partial F_l / \partial T_{\rm eff} \times \sigma(T_{\rm eff})$,
where $\sigma(T_{\rm eff}) \lesssim 100$~K and the $F_l$ derivative can
be estimated from Fig.~\ref{f7}. In any case, it 
is clear from the figure that, by neglecting any further luminosity 
correction to our data for the unsampled luminosity, we would be overestimating 
$m_{\rm bol}$ by at most 0.3~mag.}

Starting from the bolometric flux (which also includes the unsampled 
luminosity fraction, according to our procedure), the apparent magnitude
for each star derives as m$_{\rm bol}  = -2.5\,\log f_{\rm bol} + {\rm Z.P.}$.
If we {\it assume} for the Sun  an absolute M$_{\rm bol} = +4.72$, and 
$L_\odot = 3.89\,10^{33}$~erg\,s$^{-1}$, the bolometric zero point directly derives
as Z.P.~$= -11.50$~mag. On the same line, the BC scale is fixed once adopting
an observed value for the apparent $V$ magnitude of the Sun. Following \citet{lang91}, if $m_{\rm V}^\odot = -26.78$, 
then M$_{\rm V}^\odot = +4.79$ and a $BC_{\rm V}^\odot = -0.07$~mag derives.
Our output, for the whole stellar sample, is reported in col.~11 of 
Table~\ref{t11} and \ref{t12}, together with the relevant 
(dereddened) BC to the $V$ and $K$ photometric bands 
(BC$_V$ and BC$_K$, respectively in col.~12 and 13 of the tables).

\begin{figure*}
\centerline{
\psfig{file=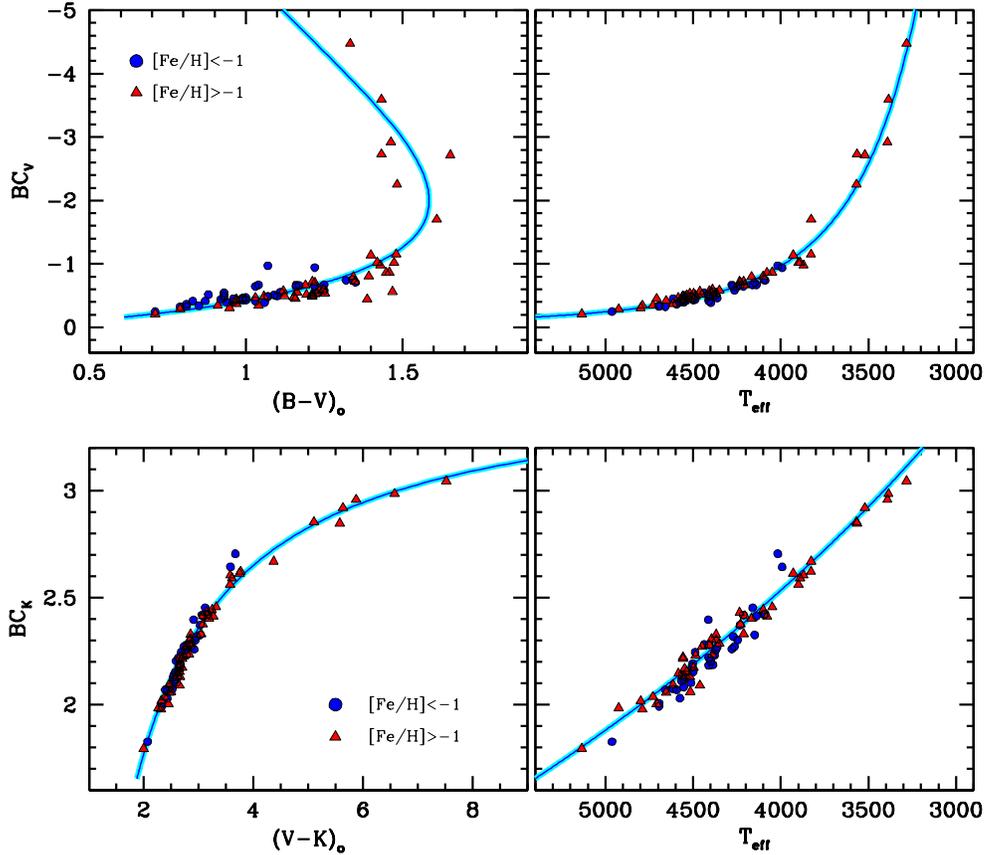,width=0.77\hsize}
}
\caption{The BC vs.\ color {\it (left panels)} and BC vs.\ T$_{\rm eff}$ {\it (right
panels)} distribution of our stellar sample (dots and triangles, for
metal-poor and metal-rich stars, respectively). Synthetic 
colors have been corrected for Galactic reddening.
Solid lines are our derived calibrations, according to the set of  
eqs.~(\ref{eq:ddt}) and (\ref{eq:ddc}).
}
\label{f9}
\end{figure*}

\section{Results and discussion}

The data of Tables~\ref{t11} and \ref{t12} are the main output of our analysis. 
According to our results, we can explore three relevant relationships, 
linking BC
with the effective temperature of stars and with two reference colors like 
$(B-V)$ and $(V-K)$. Given the temperature range of red giants, it could
be of special relevance to consider the $K$-band BC; however, for its more 
general interest, we will also include in our discussion the more standard 
case of the BC$_V$.

\subsection{BC-color-temperature relations}
 
Like for a color-color diagram, the BC vs.\ color relationship can be 
regarded as an {\it intrinsic} (i.e.\ distance-independent) feature characterizing
the stellar SED. On the corresponding theoretical side, we want also to study here 
the resulting dependence of BC on stellar effective
temperature, a relation that allows us to more directly match the 
observations with the theoretical predictions of stellar model atmospheres.

In a first set of plots (see Fig.~\ref{f9}), we display the observed distribution 
of our stars in the different planes. In order to single out any possible
dependence on chemical composition of stars, we marked differently 
metal-poor ($[Fe/H]< -1.0$~dex, dots) and metal-rich ($[Fe/H]> -1.0$~dex, 
triangles) objects. For better convenience
in our study, we also fitted the overall distribution analytically;
a useful set of fitting functions for the BC vs.\ T$_{\rm eff}$ relations 
along the $3300 \lesssim T_{\rm eff} \lesssim 5000$~K temperature range results:
\begin{equation}
\left\{
\begin{array}{ll}
{\rm BC}_{\rm V}  =  -{\rm exp}(27500/{\rm T}_{\rm eff})/1000 &   \\
\hfill  (\sigma_{\rm BC}, \rho)&= (0.11, 0.989) \\
\\ 
{\rm BC}_{\rm K} =  -6.75 \log ({\rm T}_{\rm eff}/9500) &   \\  
\hfill  (\sigma_{\rm BC}, \rho)&= (0.05, 0.978).
\end{array}
\right.
\label{eq:ddt}
\end{equation}

As for the color relations, the non-monotonic trend of BC$_V$ vs.\ $(B-V)$
(see left upper panel in Fig.~\ref{f9}) prevents us to use the color as independent 
(i.e.\ ``input'') variable in our fit. In this case we had therefore to 
adjust an inverse relation, assuming BC as the running variable. 
The corresponding set of analytical solutions, along the same temperature 
range of the previous equation set, eventually results:

\begin{equation}
\left\{
\begin{array}{ll}
{\rm B-V}  = 1.906\,\left[{\rm BC}_{\rm V}^2\,{\rm exp}({\rm BC}_{\rm V})\right]^{0.3} &  \\
\hfill (\sigma_{\rm BV}, \rho)&= (0.11, 0.863) \\ 
\\ 
{\rm V-K} = 1/(1-0.283~{\rm BC}_{\rm K}) &   \\ 
\hfill (\sigma_{\rm VK}, \rho)&= (0.13, 0.991), 
\end{array}
\right.
\label{eq:ddc}
\end{equation}
All these fits are superposed to the data of Fig.~\ref{f9} as a solid line.

Just on the basis of our data note how difficult it is to firmly constrain the 
$(B-V)$ vs.\ BC$_V$ behaviour at very low temperature. 
From one hand, in fact, the intervening effect of the TiO absoprtion at 
visual wavelength \citep{kucinskas05} makes the $(B-V)$ color of stars
cooler than $\sim 3700$~K to strongly saturate reaching a maximum of about 
$(B-V)_{\rm max} \simeq 1.5$ and turning back to bluer values for later M-type
stars. On the other hand, the apparent trend of our sample in this range is evidently 
biased by the NGC~6791 stellar population with just a few super metal rich 
giants constraining the BC$_V$ trend at the most extreme negative values.

\begin{figure}
\centerline{
\psfig{file=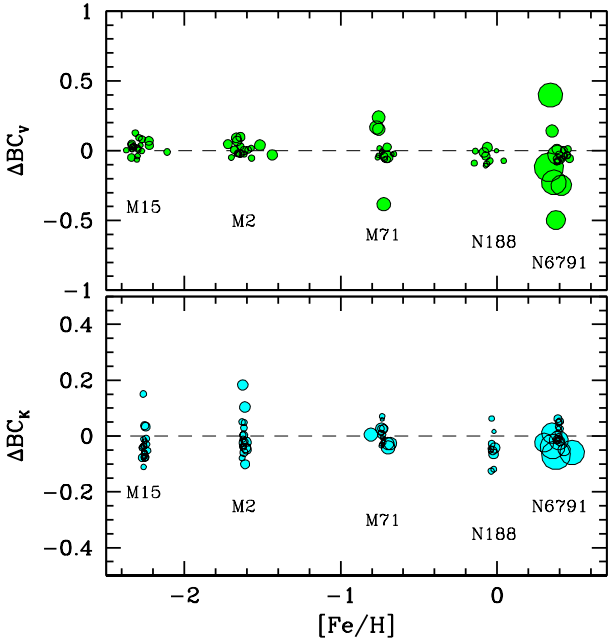,width=\hsize}
}
\caption{The distribution of BC residuals for our stellar sample vs.\
cluster metallicity. The displayed $\Delta$BC is intended
as the difference between the values of cols.\ 12 and 13 of Tables~\ref{t11}
and \ref{t12} and the output of eq.~(\ref{eq:ddt}) entering with the adopted
effective temperature of stars, as in col.\ 10 of the tables. Note the lack of
any evident correlation with [Fe/H], as discussed in more detail in Sec.~5.2.
Data in the plot have been slightly spread around the cluster [Fe/H] value 
for better reading. Dot size is inversely proportional to star temperature
(i.e.\ bigger dots = cooler red giants).
}
\label{f10}
\end{figure}

\subsection{BC response to metallicity}

As a part of our observing strategy, the sampled stellar population of
the five clusters would in principle allow to better single out any
possible dependence of BC on stellar chemical composition. As far as Helium
content is concerned, for instance, this problem has already been tackled 
by \citet{girardi07} through a series of theoretical models
based on the \citet{kurucz92} {\sc Atlas9} model atmospheres. As a main result
of their discussion, these authors did not find any relevant impact on 
stellar BC to optical photometric bands when Helium 
changes up to $\Delta Y = +0.2$, {\it for fixed effective temperature}.
To some extent, this is a not so surprising behaviour; Helium is in fact a
substantial contributor to mean particle weight of stellar plasma but
a negligible contributor to chemical opacity. Accordingly, with 
varying $Y$ in the chemical mix, one has to expect
a much more explicit impact on stellar temperature {\it for fixed mass of stars},
rather than on colors or SED {\it for fixed effective temperature}
(as explored by \citealp{girardi07} models, indeed). 

The situation might in principle be different for the metals, mainly
through their pervasive effect on stellar blanketing at short
wavelength. In addition, metals are the basic ingredients required to produce
molecules like TiO, SiH or CH, whose impact may be extremely relevant
at blue and visual wavelength, when effective temperature lowers below 3500~K
\citep{kucinskas05,bertone08}.

\begin{figure}
\centerline{
\psfig{file=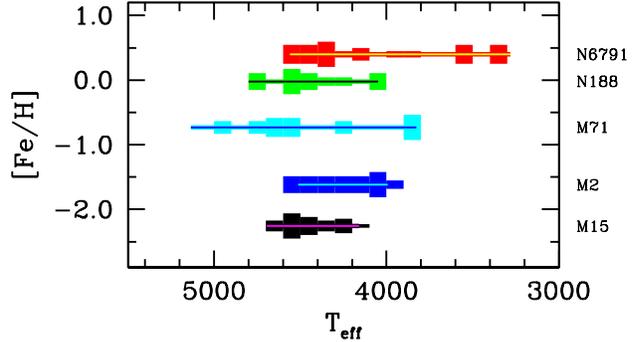,width=\hsize}
}
\caption{Temperature distribution of red-giant stars in each of the five clusters 
of our sample, according to Table~\ref{t11} and \ref{t12}. Line tickness
is proportional to the star density along the spanned temperature range.
Note that only cluster NGC~6791 contains stars cooler than $\sim 3800$~K.
}
\label{f11b}
\end{figure}

Taking the results of Tables \ref{t11} and \ref{t12} as a reference, in Fig.~\ref{f10} 
we plot the BC residual distribution computed as a difference between the 
inferred BC (cols.\ 12 and 13 in the tables)
and the ``mean'' locus of eq.~(\ref{eq:ddt}), once entering the equations 
with the fiducial $\langle$T$\rangle$ of col.~10. 
The BC residuals are displayed  along the [Fe/H] distribution of the five
star clusters, as labelled on the plots. Just a glance to both panels
of the figure makes evident the lack of any drift of BC
with stellar metallicity. Within the accuracy limits of our analysis,
this means that two red giant stars of the same effective temperature 
but different [Fe/H] have virtually indistiguishable values of BC to $V$ 
and $K$ bands.

On the other hand, to correctly understand our conclusion, one has to
pay attention to the different temperature regimes that mark spectral properties 
of red-giant stars. In fact, stars warmer than $\sim 4000$~K may have
their SED depressed at short wavelength mostly in force of {\it atomic} transitions
of Fe and other metals; on the contrary, for a cooler temperature, the metal 
opacity mainly acts in the form of {\it molecular} absorptions,
making the broad band systems the prevailing features that modulate the stellar 
SED. As a consequence, while for stars of spectral type G or earlier any change of
$Z$ simply implies a change in the blanketing strength, this may not straightforwardly
be the case for later spectral types, where molecules play a much more 
entangled role with changing $T_{\rm eff}$.

In order to better quantify the terms of our analysis, in this respect, we 
display in Fig.~\ref{f11b} the temperature distribution of stars in our sample 
across the metallicity range spanned by the five clusters considered. 
As a stricking feature, note that only
for NGC~6791 we are able to probe stars cooler than $\sim 3800$~K.
The obvious {\it caveat} in our discussion is therefore that we can only
assess the impact of {\it atomic} blanketing on stellar BC, while no firm
conclusions can be drawn for the BC dependence on molecular absorption,
facing the evident bias of our star sample against cool 
($T_{\rm eff} \ll 4000$) objects.

\begin{figure}
\centerline{
\psfig{file=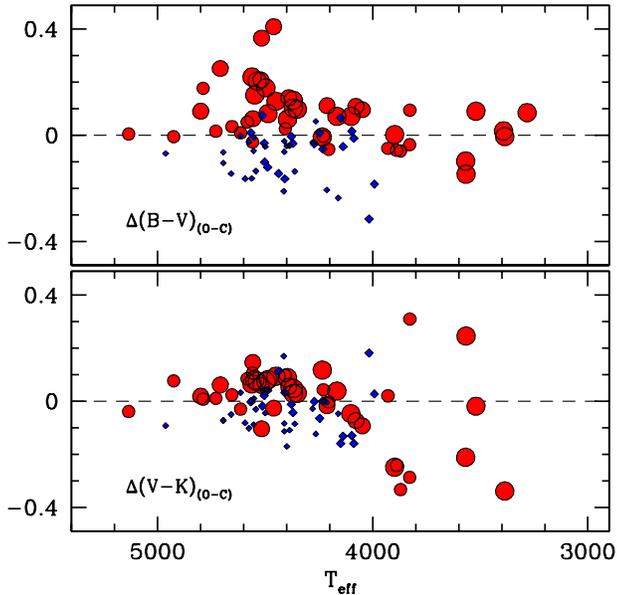,width=\hsize}
}
\caption{Residual $(B-V)$ and $(V-K)$ distribution vs.\ stellar temperature for stars
in Table~\ref{t11} and \ref{t12}. Color residual is computed as a difference between 
observed and expected values by entering eq.~(\ref{eq:ddc}) with the BC output
of eq.~(\ref{eq:ddt}). Metal-poor and metal-rich stars are singled out by diamonds
and dot markers, respectively, taking the value $[Fe/H] = -1.0$~dex as a reference 
threshold. Marker size increases with [Fe/H], throughout.
}
\label{f11}
\end{figure}
  
As far as the blanketing is the prevailing mechanism at work in G-K stars,
basic physics of stellar atmospheres leads to conclude that the $V$-band (and even 
more the $K$-band) luminosity are nearly unaffected by metal absorption, so that 
BC cannot vary much with [Fe/H]. 
Rather, $B$ (and even more $U$) magnitudes must be more strongly modulated
by metal abundance making BC$_B$ (and BC$_U$) more directly sensitive
to [Fe/H]. On the other hand, as $BC_B = BC_V -(B-V)$,
one can straight ``translate'' this metallicity effect in terms of apparent 
$(B-V)$ color change. This is shown in Fig.~\ref{f11}, where for each star in
our sample we computed the
residual $(B-V)$ and $(V-K)$ color as a difference between observed and 
expected values by entering eq.~(\ref{eq:ddc}) with the fitted value of
BC as from eq.~(\ref{eq:ddt}). Metallicity is traced in the plot by the
marker size (the bigger the marker the higher the [Fe/H] value); again, we 
discriminate between metal-poor (diamonds) and metal-rich (dots) stars,
taking the value $[Fe/H] = -1.0$~dex as a reference threshold. 

A trend of $\Delta (B-V)$ vs.\ cluster metallicity is now clearly evident, 
with the metal-poor and metal-rich star samples neatly segregated in the plot, the
latter stars displaying a ``redder'' $(B-V)$ color (and correspondingly
a positive color residual) {\it for fixed effective temperature}.
On the contrary, note that both ``metal-poor'' and ``metal-rich'' stars are
well mixed in the $\Delta (V-K)$ plot, witnessing once more the 
property of the $V-K$ color as a virtually metal-independent feature.

Considering in more detail the $\Delta (B-V)$ distribution vs.\ cluster metallicity,
a fit to the data provides:\footnote{Of course, following our previous 
arguments, we had to exclude from our analysis cluster NGC~6791, for its 
obvious bias in contraining the emipirical T$_{\rm eff}$ vs. $(B-V)$ 
relationship for stars at supersolar metallicity.}
\begin{equation}
\begin{array}{rrlr}
-\Delta BC_B \equiv \Delta (B-V) = & 0.10& [Fe/H] & +0.13 \\
                                  & \pm1 &        & \pm2 \\
\end{array}
\label{eq:bcb}
\end{equation}
with error bars at $1\sigma$ level and $({\rm rms}, \rho) = (0.09\, {\rm mag}, 0.70)$.

\begin{figure*}
\centerline{
\psfig{file=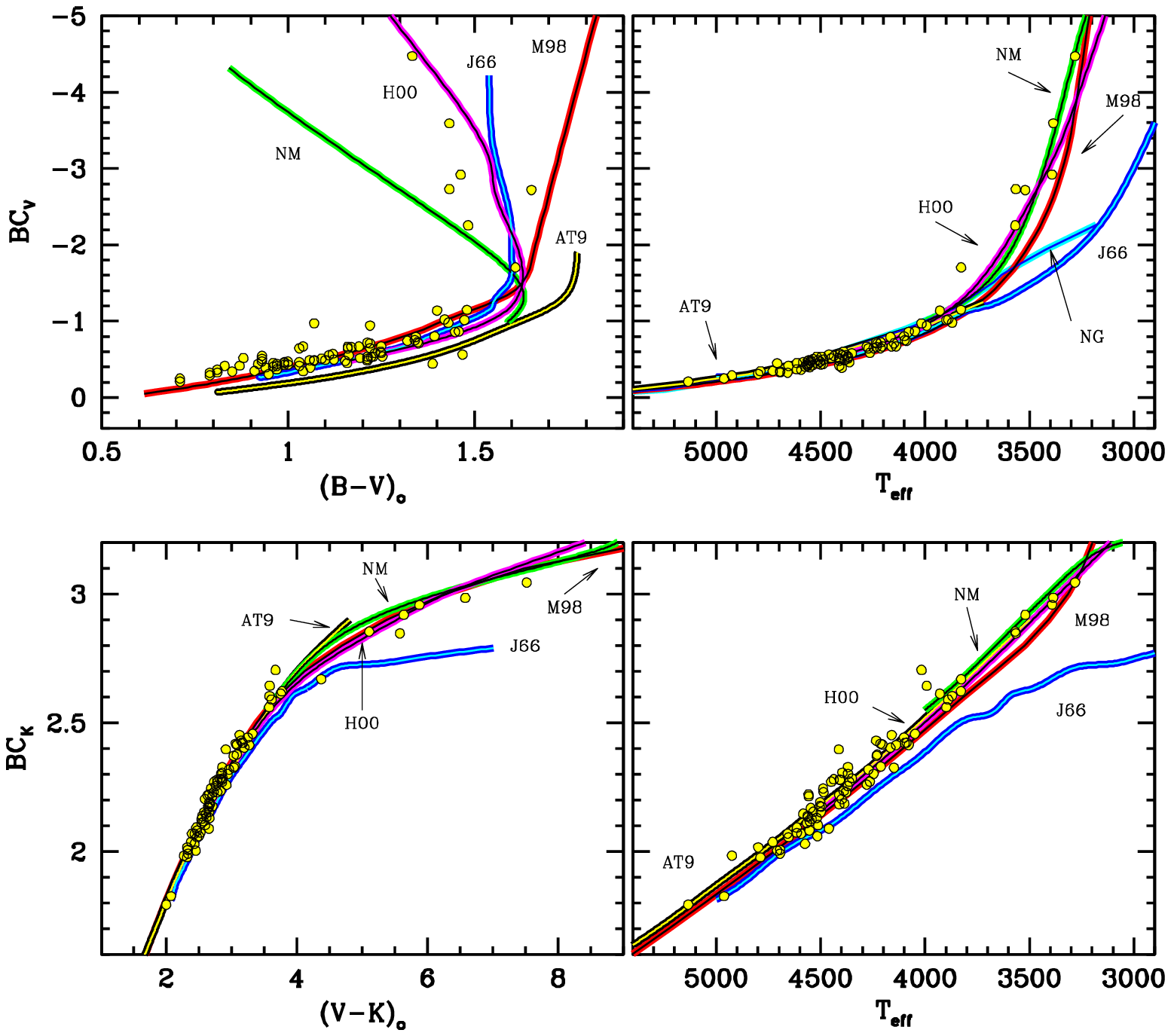,width=0.94\hsize}
}
\caption{Same as Fig.~\ref{f9}, but comparing our data with different theoretical
and empirical calibrations from \citet[][ ``J66'' labels]{johnson66}, \citet{bertone04}
using {\sc Atlas9} \citep[][``AT9'']{kurucz92} and {\sc NextGen} \citet[][``NG'']{hauschildt99}
synthesis codes for model atmosphere computation, \citet[][``M98'']{montegriffo98},
\citet[][``H00'']{houd00} using {\sc Marcs} theoretical code by 
\citep{bell78}, and its updated version ({\sc NMarcs}), as in 
\citet{plez92} and \citet[][``NM'']{bessell98}.
}
\label{f12}
\end{figure*}

\subsection{Comparison with other BC scales}

For a better understanding of our results it is relevant to compare
our output with other popular calibration scales often taken as a
reference in the current literature and especially attempting to extend their
analysis to cool (T$_{\rm eff} \la 3500$~K) stellar temperatures.
In particular, we will focus here on different theoretical BC calibrations 
relying on the three leading codes for advanced computation of
stellar model atmospheres, namely {\sc Atlas9} 
\citep[][hereafter labelled as ``AT9'']{kurucz92}, {\sc Nextgen} 
\citep[][``NG'']{hauschildt99}, both as reported by \citet{bertone04}, and {\sc Marcs} 
\citep[][as adopted by Houdashelt et al. 2000, ``H00' label]{bell78}
also in its updated versions ({\sc NMarcs}, as in \citet{plez92}
and \citet[][``NM'']{bessell98}.

We will also consider in our analysis two empirical studies, i.e.\ the ones 
of \citet[][referred to as ``J66'']{johnson66} and 
\citet[][labelled as ``M98'']{montegriffo98}, both based on a careful 
analysis of infrared colors to assess the problem of the bolometric 
correction and a self-consistent temperature scale for red giant stars. 
All the bolometric scales in the figure have been shifted such as
to agree with our assumption that $BC_{\rm V}^\odot = -0.07$~mag.

A synoptic look of the different theoretical and empirical frameworks
is eased by the four panels of Fig.~\ref{f12}, where we report the BC$_V$ and
BC$_K$ scales vs.\ observables (i.e.\ $(B-V)$ and $(V-K)$ colors, respectively)
and theoretical (T$_{\rm eff}$) reference quantities.
In all respects, this figure is fully equivalent with, and can be compared 
to, Fig.~\ref{f9}, where we reported our own results.

Just a quick look to the different curves of Fig.~\ref{f12} gives an immediate
picture of the inherent uncertainties in predicted BC according to the different 
calibration scales. The big issue, in this regard, much deals with the way 
models can reproduce cool stars and observations can account for the $(B-V)$ 
``saturation'' vs.\ temperature consequent to the shifted emission toward 
longer wavebands when stars become cooler than 3500~K. 
This effect makes the $B$-luminosity contribution to drop to nominal 
values among red giants, and the increasingly important role of molecular 
absorption strongly modulates optical colors of K- and M-type stars.

The still inadequate theoretical performance in modelling such cool
stars with convenient accuracy fatally frustrates also any empirical effort 
to derive a firm temperature scale and an accurate abundance analysis 
for stars at the extreme edge of the temperature distribution
(see, e.g., \citealt{bertone08} and \citealt{olling09}, for useful 
considerations on this subject).

As far as the BC$_V$ vs.\ $(B-V)$ behaviour is concerned, the reference
calibrations display the largest spread, with M98 predicting 
increasingly redder stars with decreasing temperature. At the opposite,
NM predicts a sharp color ``turnback'', with BC$_V$ 
increasing in absolute value among cool stars getting bluer and bluer.
Definitely, the emipirical calibration by J66 still remains a reference
one, fairly well tracking the observations. This trend is very closely
replied also by the {\sc Marcs} models by H00, that provide an even
better match to the data and a substantial agreement with our fitting 
function as in Fig.~\ref{f9}.

By converting colors to the theoretical plane of effective temperature
(right upper panel of Fig.~\ref{f12}), the picture slightly changes,
in particular with a stricking discrepancy of the J66 and the
theoretical NG temperature scale for T$_{\rm eff} \la 3800$~K. 
Both sources predict, in fact, much
shallower corrections for cool stars than we observe.
An overall agreement has to be reported, on the contrary, among the
other calibrations, all replying our eq.~(\ref{eq:ddt}).

The situation is much eased in the infrared domain, where a monotonic
relationship between $(V-K)$ color and BC$_K$ characterizes red giants
stars. In this new framework both the theoretical and empirical planes
are well reproduced by the different calibration scales, with the
only remarkable exception of J66 that, to some extent ``allows'' stars to 
store a bigger fraction of their bolometric luminosity in the infrared.
This leads to a tipping BC$_K \simeq 2.7$ and a too ``red'' $(V-K)$
for a given value of T$_{\rm eff}$.

Combining the different pieces of information coming from these
comparisons, it seems that the H00 {\sc Marcs} models are by far
the best ones in matching our BC estimates, closely replying
in every panel of Fig.~\ref{f12} our empirical fitting functions of 
eqs.~(\ref{eq:ddt}), (\ref{eq:ddc}) and Fig.~\ref{f9}.
In spite of this comforting appearence, however, this conclusion may
be even more puzzling from a physical point of view, as the H00 
models have been {\it a fortiori} tuned up such as to reproduce
the observed colors of M stars. As described by the authors, this
required in particular to strongly enhance the assumed TiO opacities
well beyond the admitted physical range suggested by molecular theory
and implemented in the ``standard'' {\sc Marcs} library \citep{gustaffson08}.

\section{Summary and conclusions} 

The firm knowledge of a fully reliable link between observations and 
stellar evolution models is a basic, crucial requirement for any safe 
use of stellar clocks and population synthesis templates in the study 
and interpretation of the integrated spectrophotometric properties of 
distant galaxies. Actually, the ``stellar path'' to cosmology is strictly
dependent, among others, on the accurate determination of 
the bolometric emission of stars, with varying effective temperatures 
and chemical abundance.

In this framework, we have tackled the central question of the possible  
BC dependence on stellar metallicity by securing spectroscopic observations 
for a wide sample of 92 red-giant stars in five
(3 globular + 2 open) Galactic clusters along the full metallicity range from
$[Fe/H] = -2.2$ up to +0.4 (see Sec.~3). Spectra cover the   
wavelength range from 3500~\AA\ to $2.5\mu$m, collecting optical
and IR observations. A delicate task for the final settlement of our stellar 
database dealt with the accurate flux calibration and a consistent
match of the optical and near-IR sides of the spectra such as to
reproduce, for each star, the broad-band $BVR_cI_cJHK$ 
photometry available in the literature (Sec.~2 and 3). Allover, we
are confident that stellar SED along the entire sampled wavelength
range has been set up within a $\pm 10$\% internal accuracy (see Table~\ref{t7}
and Fig.~\ref{f2}).

According to our previous arguments, however, one has also to carefully 
account for the lost contribution of ultraviolet and far-IR luminosity to 
the bolometric flux, depending on the effective temperature 
of stars. Based on the \citet{alonso99} T$_{\rm eff}$-color fitting functions, 
we took the four colors $(B-V)$, $(J-K)$, $(V-I_c)$, and $(V-K)$ as a reference
for our calibration, leading to constrain T$_{\rm eff}$ for each 
stars in our sample within an estimated error better than $\pm 100$~K 
(see Sec.~4.1), along the whole spanned temperature range ($3300 \le 
T_{\rm eff} \le 5000$~K).

The fiducial temperature allowed us to shape the unsampled portion of the SED at 
UV and far-IR wavelength by assuming a black-body emission independently
rescaled such as to connect the short and long wavelength edge of the observed
spectra. As shown in Sec.~4.2 (see also Fig.~\ref{f7}), under the 
black-body assumption, the internal uncertainty in our temperature scale 
only impact by a few 0.01~mag uncertainty in the inferred bolometric magnitude 
of our stars. In any case, by fully neglecting any unsampled spectral contribution, 
our data would be overestimating M$_{\rm bol}$ by at most 0.3 mag.

Making use of our new database, we have been able to draw a convenient set of
fitting functions for the BC vs.\ T$_{\rm eff}$, valid over the interval 
$ 3300 \le {\rm T}_{\rm eff} \le 5000$~K (see Sec.~5.1, eq.~\ref{eq:ddt}). 
Similar relationships for BC vs.\ stellar colors cannot be straightforwardly
derived (eq.~\ref{eq:ddc}), especially for the $(B-V)$, which shows a strong 
saturation effect for stars cooler than 3700~K, in consequence of the intervening 
TiO absorption at visual wavelength \citep{kucinskas05}.
In assessing properties of such very cool stars, however, one has also to 
consider that our sample is strongly biased against high-metallicity values 
as only the red giant branch of NGC~6791 ($[Fe/H] = +0.4$) hosts stars 
with T$_{\rm eff} < 3700$~K.

Thanks to the wide [Fe/H] range spanned by G stars in the five clusters considered 
here, we explored the possible BC dependence on stellar metallicity. As far as 
atomic transitions prevail as the main source of metal opacity in the spectra of 
relatively warm (T$_{\rm eff} \gtrsim 4000$~K) stars, our data confirm that 
no evident trend of BC with [Fe/H] is in place (see Fig.~\ref{f10}). In other
words, two red giant stars of the same effective temperature but different [Fe/H] are
virtually indistiguishable in the values of BC to $V$ and $K$ bands.
Things may be different, however, for the $B$ (and even more for $U$) magnitudes, 
where the blanketing effects are more and more severe. 
In fact, Fig.~\ref{f11} clearly shows that metal-poor stars display a ``bluer''
$(B-V)$ compared to corresponding metal-rich objects with the same T$_{\rm eff}$.
This leads us to conclude that a drift may be expected for $BC_B$ such as
$BC_B \propto -0.10\,[Fe/H]$ among stars with fixed value of T$_{\rm eff}$

To consistently verify our calibrations, we have shown in Fig.~\ref{f12} plots 
of BC$_V$ and BC$_K$ vs.\ colors and T$_{\rm eff}$, respectively, by 
comparing with different theoretical and empirical calibrations currently 
available in the literature. As far as theoretical predictions are concerned, 
it seems that the H00 models are the best ones matching our data in every 
relationship. This feature is not a surprising one, however, given the 
recognized intention of the H00 calculations to match M stars via 
{\it ``ad hoc''} tuning of molecular opacity. Actually, this successful 
comparison may add a further piece of evidence, all the way, to the persisting 
limit for theory to independently assess the modelling of cool stars.

\section*{Acknowledgments}
We thank Paolo Montegriffo and Livia Origlia for useful advises and contributions
to the discussion of our results. The anonymous referee is also acknowledged for
a very careful review of the paper, and for illuminating suggestions about the 
main focus of the paper.

With pleasure we also acknowledge partial financial support of the Italian INAF 
through grant PRIN07-1.06.10.04 and the Fundaci\'on G. Galilei of 
La Palma. This work made use of the {\sc SIMBAD} database, operated by {\sc CDS}, 
Strasbourg, France, and of the data products from the Two Micron All Sky Survey, 
which is a joint project of UMASS and the CALTECH Infrared 
Processing and Analysis Center, USA, funded by NASA and NSF.

\label{lastpage}
\bsp

\begin{thebibliography}{}
\bibitem[\protect\citeauthoryear{Alonso, Arribas, \& Mart{\'{\i}}nez-Roger}{1999}]{alonso99} Alonso A., Arribas S., Mart{\'{\i}}nez-Roger C., 1999, A\&AS, 140, 261 
\bibitem[\protect\citeauthoryear{Baade}{1928}]{baade28} Baade W., 1928, AN, 232, 65 
\bibitem[\protect\citeauthoryear{Bell \& Gustafsson}{1978}]{bell78} Bell R.~A., Gustafsson B., 1978, A\&AS, 34, 229 
\bibitem[\protect\citeauthoryear{Bertone et al.}{2004}]{bertone04} Bertone E., Buzzoni A., Ch{\'a}vez M., Rodr{\'{\i}}guez-Merino L.~H., 2004, AJ, 128, 829 
\bibitem[\protect\citeauthoryear{Bertone et al.}{2008}]{bertone08} Bertone E., Buzzoni A., Ch{\'a}vez M., Rodr{\'{\i}}guez-Merino L.~H., 2008, A\&A, 485, 823 
\bibitem[\protect\citeauthoryear{Bessell}{1979}]{bessell79} Bessell M.~S., 1979, PASP, 91, 589 
\bibitem[\protect\citeauthoryear{Bessell \& Wood}{1984}]{bessell84} Bessell M.~S., Wood P.~R., 1984, PASP, 96, 247 
\bibitem[\protect\citeauthoryear{Bessell, Castelli, \& Plez}{1998}]{bessell98} Bessell M.~S., Castelli F., Plez B., 1998, A\&A, 333, 231 
\bibitem[\protect\citeauthoryear{Blackwell \& Lynas-Gray}{1994}]{black94} Blackwell D.~E., Lynas-Gray A.~E., 1994, A\&A, 282, 899 
\bibitem[\protect\citeauthoryear{Blackwell, Shallis, \& Selby}{1979}]{black79} Blackwell D.~E., Shallis M.~J., Selby M.~J., 1979, MNRAS, 188, 847 
\bibitem[\protect\citeauthoryear{Blackwell, Petford, \& Shallis}{1980}]{black80} Blackwell D.~E., Petford A.~D., Shallis M.~J., 1980, A\&A, 82, 249 
\bibitem[\protect\citeauthoryear{Buzzoni}{1989}]{buzzoni89} Buzzoni A., 1989, ApJS, 71, 817 
\bibitem[\protect\citeauthoryear{Buzzoni}{2005}]{buzzoni05} Buzzoni A., 2005, MNRAS, 361, 725 
\bibitem[\protect\citeauthoryear{Cassisi et al.}{2004}]{cassisi04} Cassisi S., Salaris M., Castelli F., Pietrinferni A., 2004, ApJ, 616, 498 
\bibitem[\protect\citeauthoryear{Code et al.}{1976}]{code76} Code A.~D., Bless R.~C., Davis J., Brown R.~H., 1976, ApJ, 203, 417 
\bibitem[\protect\citeauthoryear{Cohen, Briley, \& Stetson}{2005}]{cohen05} Cohen J.~G., Briley M.~M., Stetson P.~B., 2005, AJ, 130, 1177 
\bibitem[\protect\citeauthoryear{de Marchi et al.}{2007}]{demarchi07} de Marchi F., et al., 2007, A\&A, 471, 515 
\bibitem[\protect\citeauthoryear{di Benedetto \& Rabbia}{1987}]{dibenedetto87} di Benedetto G.~P., Rabbia Y., 1987, A\&A, 188, 114 
\bibitem[\protect\citeauthoryear{Dyck et al.}{1996}]{dyck96} Dyck H.~M., Benson J.~A., van Belle G.~T., Ridgway S.~T., 1996, AJ, 111, 1705 
\bibitem[\protect\citeauthoryear{Flower}{1975}]{flower75} Flower P.~J., 1975, A\&A, 41, 391 
\bibitem[\protect\citeauthoryear{Flower}{1977}]{flower77} Flower P.~J., 1977, A\&A, 54, 31 
\bibitem[\protect\citeauthoryear{Flower}{1996}]{flower96} Flower P.~J., 1996, ApJ, 469, 355 
\bibitem[\protect\citeauthoryear{Frogel et al.}{1978}]{frogel78} Frogel J.~A., Persson S.~E., Matthews K., Aaronson M., 1978, ApJ, 220, 75 
\bibitem[\protect\citeauthoryear{Fuensalida \& Alonso}{1998}]{fuensalida98} Fuensalida J.~J., Alonso A., 1998, NewAR, 42, 543 
\bibitem[\protect\citeauthoryear{Geffert \& Maintz}{2000}]{geffert00} Geffert M., Maintz G., 2000, A\&AS, 144, 227 
\bibitem[\protect\citeauthoryear{Girardi et al.}{2007}]{girardi07} Girardi L., Castelli F., Bertelli G., Nasi E., 2007, A\&A, 468, 657 
\bibitem[\protect\citeauthoryear{Gonz{\'a}lez Hern{\'a}ndez \& Bonifacio}{2009}]{gonzalez09} Gonz{\'a}lez Hern{\'a}ndez J.~I., Bonifacio P., 2009, A\&A, 497, 497 
\bibitem[\protect\citeauthoryear{Gratton et al.}{1982}]{gratton82} Gratton L., Gaudenzi S., Rossi C., Gratton R.~G., 1982, MNRAS, 201, 807 
\bibitem[\protect\citeauthoryear{Gustafsson et al.}{2008}]{gustaffson08} Gustafsson B., Edvardsson B., Eriksson K., J{\o}rgensen U.~G., Nordlund {\AA}., Plez B., 2008, A\&A, 486, 951 
\bibitem[\protect\citeauthoryear{Hauschildt, Allard, \& Baron}{1999}]{hauschildt99} Hauschildt P.~H., Allard F., Baron E., 1999, ApJ, 512, 377 
\bibitem[\protect\citeauthoryear{Houdashelt, Bell, \& Sweigart}{2000}]{houd00} Houdashelt M.~L., Bell R.~A., Sweigart A.~V., 2000, AJ, 119, 1448 
\bibitem[\protect\citeauthoryear{Hunt et al.}{1998}]{hunt98} Hunt L.~K., Mannucci F., Testi L., Migliorini S., Stanga R.~M., Baffa C., Lisi F., Vanzi L., 1998, AJ, 115, 2594 
\bibitem[\protect\citeauthoryear{Johnson}{1966}]{johnson66} Johnson H.~L., 1966, ARA\&A, 4, 193 
\bibitem[\protect\citeauthoryear{Kaluzny \& Rucinski}{1995}]{kaluzny95} Kaluzny J., Rucinski S.~M., 1995, A\&AS, 114, 1 
\bibitem[\protect\citeauthoryear{Kuiper}{1938}]{kuiper38} Kuiper G.~P., 1938, ApJ, 88, 429 
\bibitem[\protect\citeauthoryear{Ku{\v c}inskas et al.}{2005}]{kucinskas05} Ku{\v c}inskas A., Hauschildt P.~H., Ludwig H.-G., Brott I., Vansevi{\v c}ius V., Lindegren L., Tanab{\'e} T., Allard F., 2005, A\&A, 442, 281 
\bibitem[\protect\citeauthoryear{Kurucz}{1992}]{kurucz92} Kurucz R.~L., 1992, IAUS, 149, 225 
\bibitem[\protect\citeauthoryear{Lan{\c c}on \& Mouhcine}{2002}]{lancon02} Lan{\c c}on A., Mouhcine M., 2002, A\&A, 393, 167 
\bibitem[\protect\citeauthoryear{Lang}{1991}]{lang91} Lang K.R., 1991, Astrophysical Data: planets and stars (Heidelberg: Springer Verlag)
\bibitem[\protect\citeauthoryear{Manduca \& Bell}{1979}]{manduca79} Manduca A., Bell R.~A., 1979, PASP, 91, 848 
\bibitem[\protect\citeauthoryear{Massey et al.}{1988}]{massey88} Massey P., Strobel K., Barnes J.~V., Anderson E., 1988, ApJ, 328, 315 
\bibitem[\protect\citeauthoryear{Mochejska, Stanek, \& Kaluzny}{2003}]{moc03} Mochejska B.~J., Stanek K.~Z., Kaluzny J., 2003, AJ, 125, 3175 
\bibitem[\protect\citeauthoryear{Montegriffo et al.}{1998}]{montegriffo98} Montegriffo P., Ferraro F.~R., Origlia L., Fusi Pecci F., 1998, MNRAS, 297, 872 
\bibitem[\protect\citeauthoryear{Olling et al.}{2009}]{olling09} Olling R.~P., et al., 2009, Astro2010: The Astronomy and Astrophysics Decadal Survey, Science White Papers, no. 226 (see also {\sf astroph/0902.3197}) 
\bibitem[\protect\citeauthoryear{Perrin et al.}{1998}]{perrin98} Perrin G., Coude Du Foresto V., Ridgway S.~T., Mariotti J.-M., Traub W.~A., Carleton N.~P., Lacasse M.~G., 1998, A\&A, 331, 619 
\bibitem[\protect\citeauthoryear{Platais et al.}{2003}]{platais03} Platais I., Kozhurina-Platais V., Mathieu R.~D., Girard T.~M., van Altena W.~F., 2003, AJ, 126, 2922 
\bibitem[\protect\citeauthoryear{Plez, Brett, \& Nordlund}{1992}]{plez92} Plez B., Brett J.~M., Nordlund A., 1992, A\&A, 256, 551 
\bibitem[\protect\citeauthoryear{Ram{\'{\i}}rez \& Mel{\'e}ndez}{2005}]{ramirez05} Ram{\'{\i}}rez I., Mel{\'e}ndez J., 2005, ApJ, 626, 446 
\bibitem[\protect\citeauthoryear{Richichi et al.}{1998}]{richichi98} Richichi A., Ragland S., Stecklum B., Leinert C., 1998, A\&A, 338, 527 
\bibitem[\protect\citeauthoryear{Ridgway et al.}{1980}]{ridgway80} Ridgway S.~T., Jacoby G.~H., Joyce R.~R., Wells D.~C., 1980, AJ, 85, 1496 
\bibitem[\protect\citeauthoryear{Rosenberg et al.}{2000}]{rosenberg00} Rosenberg A., Piotto G., Saviane I., Aparicio A., 2000, A\&AS, 144, 5 
\bibitem[\protect\citeauthoryear{Scheffler}{2006}]{landolt06} Scheffler, H., 2006, in Landolt-B\"{o}rnstein - Group VI Astronomy and Astrophysics, Chap. 7 (Heidelberg: Springer Verlag), p.46
\bibitem[\protect\citeauthoryear{Skrutskie et al.}{2006}]{2mass} Skrutskie M.~F., et al., 2006, AJ, 131, 1163 
\bibitem[\protect\citeauthoryear{Stetson, Bruntt, \& Grundahl}{2003}]{stetson03} Stetson P.~B., Bruntt H., Grundahl F., 2003, PASP, 115, 413 
\bibitem[\protect\citeauthoryear{Stetson, McClure, \& VandenBerg}{2004}]{stetson04} Stetson P.~B., McClure R.~D., VandenBerg D.~A., 2004, PASP, 116, 1012 
\bibitem[\protect\citeauthoryear{Tokunaga \& Vacca}{2005}]{tokunaga05} Tokunaga A.~T., Vacca W.~D., 2005, PASP, 117, 421 
\bibitem[\protect\citeauthoryear{Tripicco \& Bell}{1995}]{tripicco95} Tripicco M.~J., Bell R.~A., 1995, AJ, 110, 3035 
\bibitem[\protect\citeauthoryear{VandenBerg \& Clem}{2003}]{vdb03} VandenBerg D.~A., Clem J.~L., 2003, AJ, 126, 778 
\bibitem[\protect\citeauthoryear{Wesselink}{1969}]{wesselink69} Wesselink A.~J., 1969, MNRAS, 144, 297 
\bibitem[\protect\citeauthoryear{Worthey \& Lee}{2006}]{worthey06} Worthey G., Lee H., 2006, {\sf astro-ph/0604590} 

\end{thebibliography}
\end{document}